\documentclass[fleqn,usenatbib]{mnras}

\usepackage{newtxtext,newtxmath}
\usepackage[T1]{fontenc}
\usepackage{ae,aecompl}

\usepackage{graphicx}	
\usepackage{amssymb}	
\usepackage{amsmath}	
\usepackage{xspace}
\usepackage{breakurl}
 
\newcommand{\K}{\hbox{\it K}\xspace}		
\newcommand{\Ks}{\hbox{\it Ks}\xspace}		
\newcommand{\kms}{\hbox{km\,s$^{-1}$}\xspace} 
\newcommand{\Davies}{\hbox{{D10}}\xspace}	
\newcommand{\hh}{$\rm{H}_2$\xspace}
\newcommand{\tex}{$T_{\rm{exc}}$\xspace}
\newcommand{\ntot}{$N_{\rm{tot}}$\xspace}
\newcommand{\Ak}{$A_{\rm{K}}$\xspace}

\newcommand{\nai}{$\rm{Na\,\textsc{i}}$\xspace}

\newcommand{\brg}{$\rm{Br\gamma}$\xspace}
\newcommand{\msun}{$\rm{M}_\odot$\xspace}

\title[PCA Tomography of the NIFS K-band observations of the high-mass protostar W33A]{Principal component analysis tomography in near-infrared integral field spectroscopy of young stellar objects. I. Revisiting the high-mass protostar W33A}

\author[Navarete et al.]{
F. Navarete$^{1}$\thanks{E-mail: navarete@usp.br (FN)},
A. Damineli$^{1}$,
J.~E. Steiner$^{1}$\thanks{\textit{In Memoriam}},
R.~D.~Blum$^{2}$\\
  $^{1}$~Universidade de S\~ao Paulo, Instituto de Astronomia, Geof\'isica e Ci\^encias Atmosf\'ericas, Rua do Mat\~ao 1226,\\ Cidade Universit\'aria S\~ao Paulo-SP, 05508-090, Brasil \\
  $^{2}$~NSF's Optical–Infrared Astronomy Research Laboratory P.O. Box 26732, Tucson, AZ 85719, USA\\
}

\date{Accepted 2021 February 4. Received 2021 February 4; in original form 2020 November 29}

\pubyear{2021}

\begin{document}
\label{firstpage}
\pagerange{\pageref{firstpage}--\pageref{lastpage}}
\maketitle

\begin{abstract}
W33A is a well-known example of a high-mass young stellar object showing evidence of a circumstellar disc.
We revisited the $K$-band NIFS/Gemini North observations of the W33A protostar using principal components analysis tomography and additional post-processing routines.
Our results indicate the presence of a compact rotating disc based on the kinematics of the CO absorption features. The position-velocity diagram shows that the disc exhibits a rotation curve with velocities that rapidly decrease for radii larger than 0\farcs1 ($\sim$250\,AU) from the central source, suggesting a structure about four times more compact than previously reported.
We derived a dynamical mass of 10.0$^{+4.1}_{-2.2}$\,\msun for the ``disc+protostar'' system, about $\sim$33\% smaller than previously reported, but still compatible with high-mass protostar status.
A relatively compact H$_2$ wind was identified at the base of the large-scale outflow of W33A, with a mean visual extinction of $\sim$63 mag.
By taking advantage of supplementary near-infrared maps, we identified at least two other point-like objects driving extended structures in the vicinity of W33A, suggesting that multiple active protostars are located within the cloud.
The closest object (Source\,B) was also identified in the NIFS field of view as a faint point-like object at a projected distance of $\sim$7,000\,AU from W33A, powering extended \K-band continuum emission detected in the same field.
Another source (Source\,C) is driving a bipolar \hh jet aligned perpendicular to the rotation axis of W33A.
\end{abstract}




\section{Introduction}
\label{sec_intro}

In the last few decades, integral field unit (IFU) observations have become crucial to understand the kinematics of very complex extended objects over a variety of scales, from protostellar objects and protoplanetary discs (e.g. \citealt{Murakawa13,Alves19}) to galaxies \citep[][]{Kewley19}.
The powerful three-dimensional data cubes produced by IFU spectrographs are of increasingly large size, containing an enormous amount of information and noise that must be pulled from the cube by sophisticated means.
For example, a typical near-infrared data cube may have millions of pixels, and given the complexity of the spatial and spectral information, often times much of the information is hard to extract and may be simply ignored in the analysis in preference of a restricted subset of the data (line maps and their products, spectral analysis of bright regions, etc.).
Frequently, a large fraction of the data in these data cubes contains just instrumental or sky noise.
Thus, robust procedures to extract  information from the large data sets are increasingly required in astrophysics.

In this context, \citet{Steiner09} developed a method for the use of the Principal Component Analysis (PCA) Tomography in optical and near-infrared IFU data cubes.
PCA is a non-parametric technique, that is, it does not require any assumption or expectation of the physical and geometrical parameters of the data.
PCA compresses the data expressed as a large set of correlated variables (spatial and spectral pixels) in a small but optimal set of uncorrelated and orthogonal variables (i.e. the eigenvectors and their spatial projections), ordered by their relative importance to explain the initial data set.

PCA tomography has been recently used in extra-galactic studies \citep[e.g.][]{Ricci11,Menezes13,Ricci14,Ricci15,Dahmer19}, but no application to stellar astrophysics cases has been made yet. Thus, we applied this technique to the \K-band IFU observations of the well-known high-mass protostar W33A investigated by \citet{Davies10} to check if the PCA Tomography is able to recover their findings.
W33A is a well-known young stellar object (YSO) with indications of an active accretion process ongoing.
This protostar is associated with the Red MSX Source object G012.9090$-$00.2607, with a bolometric luminosity of $\log(L_{\rm bol}/L_\odot)=4.51$ \citep{Lumsden13} and located at the distance of 2.4\,kpc \citep{Immer13}.

\citet{Davies10} presented the study of this source through three-dimensional spectroscopy in the $K$-band (2.2\,{\micron}) using NIFS/Gemini North IFU observations.
Those authors reported the identification of the key-structures predicted by the disc-mediated accretion scenario \citep{Shu87}: a circumstellar disc traced by the CO features in absorption; a jet-like structure traced by the \hh emission at 2.12\,{\micron}; and the cavity of the jets close to their launching point traced by the spectro-astrometry of the \brg emission at milli-arcsecond scales.
Those authors derived the Keplerian mass of $\sim$15\,\msun for the ``protostar + disc'' system, assuming a distance of 3.8\,kpc.

Given the importance of W33A as one of the rare cases of high-mass YSOs with clear signs of active accretion, the \K-band IFU observations of the high-mass protostar W33A offer an excellent opportunity to verify the efficiency of PCA Tomography on reproducing the results reported by \citet{Davies10}.
%
We are also interested to check how the results are sensitive to the additional post-processing routines suggested by \citet{Steiner09} (e.g. differential atmospheric refraction correction, high-frequency noise filtering, and instrumental fingerprint removal; see also \citealt{Menezes14}), and if PCA Tomography can provide additional information based on the correlation of the spectral features present in the NIFS \K-band observations.

This work 
is organised as follows.
Section\,\ref{sec_obs} presents the observations, the post-processing routines of the data and the description of the PCA tomography technique. Section\,\ref{sec_results} presents the results and interpretation of the data through PCA tomography, the derivation of the physical parameters of the source and the analysis of auxiliary near-infrared maps.
Section\,\ref{sec_discussion} presents the discussion of the data and comparison with recent findings in the literature.
Our conclusions are given in Section\,\ref{sec_summary}.

\section{Observations and data processing}
\label{sec_obs}

\subsection{\texorpdfstring{$K$}{K}-band IFU observations}

    The \K-band ($\lambda$\,$\approx$\,2.2\,$\mu$m) IFU {observations} of W33A were obtained with the \textit{Near-Infrared Integral Field Spectrometer} \citep[NIFS,][]{McGregor03}.
    NIFS has a full wavelength coverage in the \K-band of about 4200\,\AA\xspace, with a linear dispersion of $\Delta\lambda_{\rm disp}$\,=\,2.13\,\AA/pixel$^{-1}$ and a nominal spectral resolving power of $\lambda/\Delta\lambda_{\rm disp}$\,=\,5160 \citep{Blum08}, or $\approx$\,58\,\kms in the velocity scale.
    {NIFS observes a {3\arcsec$\times$3\arcsec} field-of-view in a 60$\times$60 pixel$^2$ array, delivering a spatial sampling of 0\farcs050\,pixel$^{-1}$.}

   The observations were carried at the Gemini North 8-m telescope on the night of May 25th, 2008. The raw data of W33A were downloaded from the Gemini Science Archive (project GN-2008A-Q-44). The data include a set of calibration frames required for processing the science observations: a set of flat-field images taken with Quartz lamps on (flat-on) and off (flat-off); flat-field images taken with a Ronchi mask (Ronchi flats), used to correct the spatial distortion along the spectral dimension; a set of dark images, taken with the same exposure time of the science and telluric sources; and an Argon-Neon lamp spectrum (Ar-Ne) for wavelength calibration of the data.
   The {A2\,V} double star HR6798A+B was observed as the telluric standard star, used for correcting the telluric absorption features, and for performing the flux-calibration of the final data cubes.
   
   The data were processed using the usual reduction procedure based on IRAF/Gemini package.
   The telluric correction was applied to the data using a one-dimensional template spectrum extracted from the telluric standard star. The telluric \brg feature was removed from the template by fitting a Voigt profile.
   Then, the data cubes were constructed using the Gemini \texttt{NIFCUBE} procedure, by interpolating the original data into a final spatial pixel  (spaxel) size of 0\farcs050, resulting in a set of six science data cubes.
    {The final spatial resolution achieved in the observations is about 0\farcs14\,$\pm$\,0\farcs01, obtained as the mean FWHM of the telluric standard double star.}
   
    Residual telluric emission features were identified as relatively strong emission lines in all data cubes.
    Since these features often exhibit significant variability in the FOV that might affect the results of the PCA analysis, they were manually suppressed from the data by interpolating the adjacent spectral pixels in the affected regions for each spaxel of the data cube.

\subsection{Supplementary near-IR maps}

Additional $J$ ($\lambda$\,$\approx$\,1.25 \micron), H (1.64\,\micron), \hh (2.14\,\micron) and $K$ (2.22\,\micron) images of a {90\arcsec$\times$90\arcsec} FOV around the W33A protostar were obtained from
the UKIRT Widefield Infrared Survey for H$_2$ (UWISH2)\footnote{\url{http://astro.kent.ac.uk/uwish2/}} \citep{Froebrich11}. The large scale maps have a pixel scale of 0\farcs2\,pixel$^{-1}$ and a mean seeing of about 0\farcs96 at the \K-band, offering a complementary view of the stellar and nebular content in the vicinity of W33A.

\subsection{Post-processing of the IFU data} 
\label{sec_postprocess}

    The data cubes generated by the standard \texttt{IRAF} reduction procedure were further processed using a set of routines written in \texttt{IDL} (Interactive Data Language).
    The purpose of these additional procedures include $i)$ the atmospheric refraction correction and combination of the data cubes, $ii)$ the flux-calibration, and $iii)$ the high-frequency noise suppression from the data.

\subsubsection{Differential Atmospheric Refraction Correction}
    
\begin{figure}
	\includegraphics[width=\columnwidth]{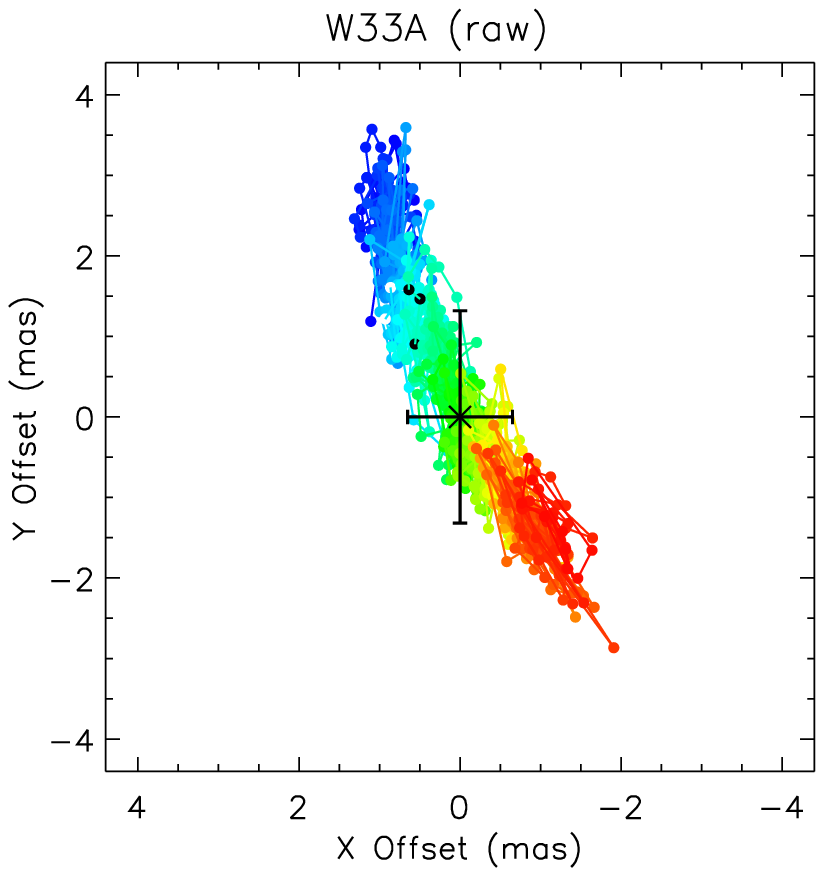}
	\includegraphics[width=\columnwidth]{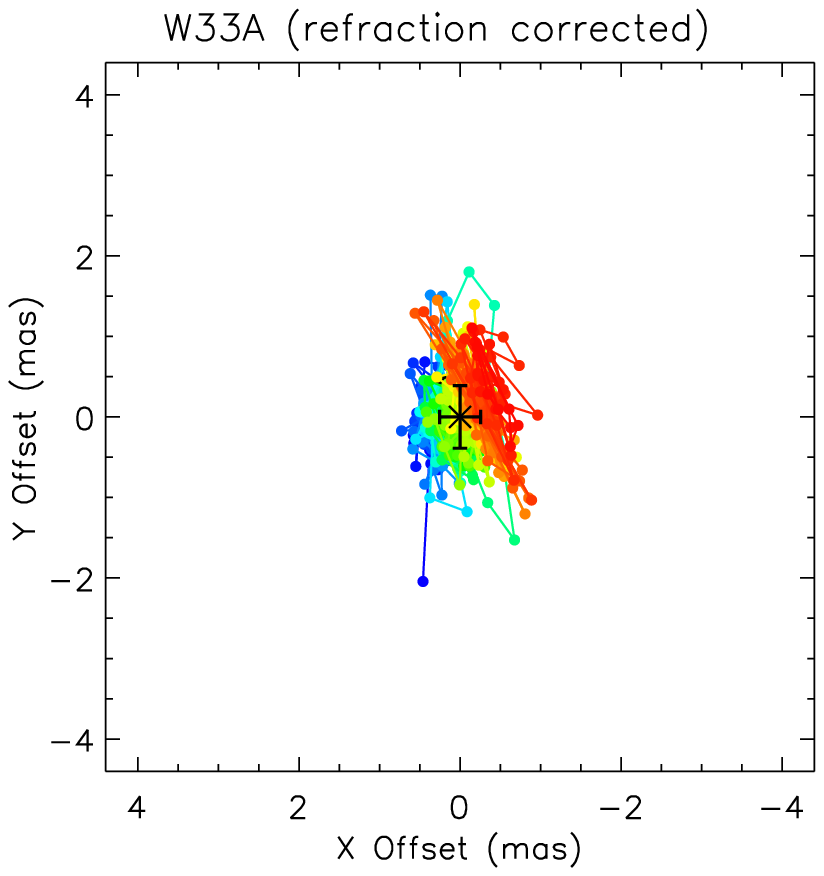} \\[-4.5ex]
    \caption{Centroid position of the W33A source as a function of the wavelength (from 2.02 to 2.42\,{\micron}, from blue to red) of the same data cube before (top panel) and after the atmospheric refraction correction (bottom). The (0,0) position is indicated as a $\times$ symbol and the error bars corresponds to the standard deviation of the position across the spectral dispersion.}
    \label{fig_atmrefraction}
\end{figure}

    The individual data cubes were corrected for the effects of the differential atmospheric refraction (DAR) using the procedure described in \citet{Menezes14}.
    Figure\,\ref{fig_atmrefraction} compares the data before and after the DAR-correction based on the centroid position of the point-like source of W33A as a function of the wavelength for the entire spectral region covered by the \K-band data cube.
    The plot shows the rms of the source position reduces from $\sim$1.5\,mas to about 0.5\,mas, indicating the DAR-correction is a significant correction to the displacement introduced by atmospheric effects at NIR wavelengths.
    The procedure adopted for performing the DAR-correction of the data cubes also takes into account any dithering offsets used for the observations, delivering a set of aligned data cubes.
The displacements introduced by the DAR effects are significant for the analysis involving the entire continuum (e.g. see Figs.\,4 and 5 from \citealt{Menezes14}), or different spectral features along the wavelength range covered by the data (e.g. the \hh lines). Indeed, \citet{Menezes14} reported that the total displacement introduced by the DAR effects can be larger than the typical NIFS spatial resolution from $J$- to \K-band wavelengths. However, we do not expect a critical enhancement of the results introduced by the DAR correction on the analysis of single spectral lines with widths of a few tens to hundreds of Angstroms. However, the single line analysis benefits from the post-processing routines aimed at the suppression of the noise from the datacube, as further presented in Sect.\,\ref{sec_butterworth} and \ref{sec_reconstruction}.
    
    A final data cube was obtained by median combining the six individual DAR-corrected data cubes.
    We compared the residuals between the median-combined and an average-combined data cube to check if the fluxes were preserved. We noted that the largest deviations were about $\sim$\,1\% of the flux in the brightest regions of the FOV, closer to the W33A point source, and $\sim$\,5\% in low signal-to-noise regions, indicating no significant discrepancy between the usage of the median or the average of the fluxes.

\subsubsection{Flux calibration of the data cubes}
\label{sec_fluxcal}

    The median-combined data cube was flux-calibrated using a custom \texttt{IDL} routine based on the \texttt{STANDARD}, \texttt{SENSFUNC} and \texttt{CALIBRATE} routines of the \texttt{IRAF}-NOAO package. The three corrections performed by each routine are applied simultaneously by computing the `calibration factor' function ($C_\lambda$), defined as:
  \begin{equation}
    C_\lambda = 2.5 \log\left( \frac{ O_\lambda }{ t_{\rm{exp}}~B_\lambda(T_{\rm{eff}},m_{Ks})} \right) + A\,E_\lambda 
    \label{eq_calibrationfunction}
  \end{equation}
\noindent where, $O_\lambda$ corresponds to the observed telluric counts of the observation (in ADU), $t_{\rm{exp}}$ is the exposure time of the observation (in seconds), $B_\lambda(T_{\rm{eff}},m_{Ks})$ is the black-body flux (in erg\,s$^{-1}$\,cm$^{-2}$\,$\mu$m$^{-1}$) at a given effective temperature $T_{\rm{eff}}$, scaled by the \Ks-band magnitude of the telluric standard, $A$ is the average airmass of the observations and $E_\lambda$ is the atmospheric extinction curve (in mag\,airmass$^{-1}$ units). 

    We adopted the \Ks-band flux of the 2MASS catalogue \citep{Skrutskie06} to derive the flux-calibration of the data cubes. Since the HR6798A+B system is unresolved at the 2MASS catalogue but resolved in the NIFS observations, the construction of $O_\lambda$ was performed by computing the integrated spectrum of {the resolved components of} the telluric standard star over the spatial pixels (spaxels) with average flux above a 3-$\sigma$ threshold above the value of the sky background.
    Then, the continuum emission was modelled by fitting a polynomial function of order $n$\,$\leq$\,4, using the outlier-resistant IDL algorithm \texttt{robust\_poly\_fit}.
    Next, $B_\lambda(T_{\rm{eff}},m_{Ks})$ was constructed by following a two-step procedure.
    First, the shape of the black-body spectrum was obtained by using the effective temperature corresponding to the spectral type of the telluric standard ({A2\,V}), taken from \citet{Tokunaga00}. 
    Then, the black-body emission was scaled using the \Ks-band flux of the telluric star {(\Ks\,=\,5.87\,mag)}.
    The 2MASS \Ks-band magnitudes ($m_{\rm Ks}$) were converted to fluxes ($F_{\rm Ks}$) using Eq\,\ref{eq_mag_flux}:
      \begin{equation}
        m_{Ks} = -2.5 \log\left( \frac{ F_{Ks} }{ F_{Ks,0} } \right)
        \label{eq_mag_flux}
      \end{equation}
\noindent where $F_{Ks,0}$\,=\,4.28\,$\cdot$\,10$^{-7}$\,erg\,s$^{-1}$\,cm$^{-2}$\,{\micron}$^{-1}$ is the zero-point flux at the \Ks-band \citep{Cohen03}.
      
    Finally, the additive term of Eq.\,\eqref{eq_calibrationfunction} requires a model for the atmospheric extinction as a function of the wavelength.
    Thus, we adopted the mean extinction through the broad \K-band filter for Mauna Kea, $E_\lambda$\,=\,0.033\,mag/airmass, reported by  \citet{Tokunaga02}.
    
    After constructing $C_\lambda$, each spaxel of the data cube (here, represented as $D_{xy\lambda}$) was divided by the exposure time of the observations, and was extinction-corrected and flux-calibrated as indicated by Eq.\,\eqref{eq_sci_fcor}:
\begin{equation}
    \log(D_{fcal,xy\lambda}) = {0.4 \cdot C_\lambda } \cdot \log\left(  \frac{ D_{xy\lambda} }{ t_{\rm{exp}} } \right)
    \label{eq_sci_fcor}
\end{equation}
    The results of the flux calibration were checked using the integrated telluric standard spectrum as follows.
    First, the total flux of the telluric standard star was integrated over the bandwidth of the 2MASS \Ks-band filter \citep[$\lambda_c$\,=\,2.159\,{\micron}, $\Delta\lambda$\,=\,0.262\,{\micron},][]{Cohen03}.
    Then, the corresponding \Ks-band magnitude was evaluated using Eq.\,\eqref{eq_mag_flux}, and compared with the 2MASS \Ks-band magnitude of the telluric star (5.87\,mag). Their difference, $\Delta K_{\rm s}$\,=\,$-$0.03\,mag, corresponds to a \Ks-band flux ratio of $\approx$1.028 between the NIFS and 2MASS fluxes, or an error of about 3 per cent.
    
\begin{figure}
	\includegraphics[width=\columnwidth]{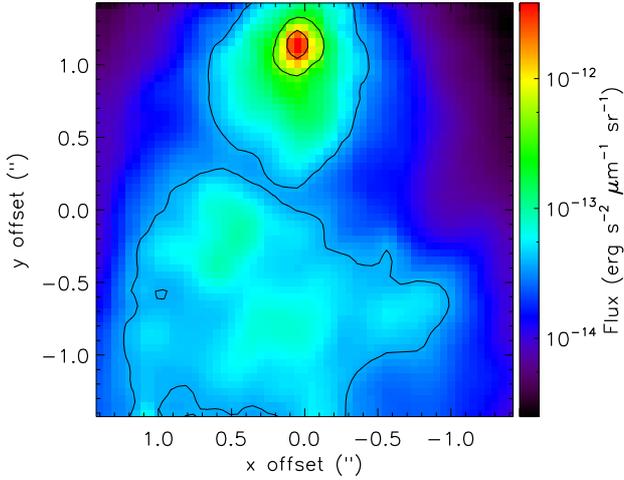} \\[-5.5ex]
    \caption{Integrated flux of the \K-band NIFS observations of W33A. The image is shown in the logarithm scale indicated by the colour bar in the right. The contours are placed at the 1, 10 an 50\% of the peak intensity of the map. The W33A point source is located around the (0\farcs0, 1\farcs2) position in the FOV.}
    \label{fig_w33a_nifKband}
\end{figure}

    Figure\,\ref{fig_w33a_nifKband} shows the integrated emission of the flux calibrated W33A data, showing the point like object at the upper central region of the FOV (0\farcs0, 1\farcs2) and the associated \K-band continuum nebulosity towards the South.

\subsubsection{High-frequency spatial noise filtering}
\label{sec_butterworth}

NIFS data are provided in a grid of $x$, $y$ and $\lambda$ coordinates, where the astrophysical information is often mixed with a complex noise structure, including the photon noise, high-frequency noise and systematic instrumental noise that is often associated with astronomical data cubes (see, e.g., \citealt{Menezes14}). Although the photon noise is a well-known and inherent source of noise in astronomical observations, the nature of the last two components are in general unknown and may be introduced during the data acquisition or data processing (e.g. the resampling of the original 0\farcs103\,$\times$\,0\farcs43 to the 0\farcs050\,$\times$\,0\farcs050 spatial pixels). Despite the unknown origin of the noise in IFU observations, a considerable fraction of it can be successfully suppressed from the data based on the application of a low-pass Butterworth filter (explained below) and the PCA Tomography (Sect.\,\ref{sec_pca}) on the data.

In the Fourier domain, the spatial information can be expressed in terms of frequencies ($\nu_{\rm x}$, $\nu_{\rm y}$), allowing us to identify and suppress  high-frequency noise using an appropriate low-pass band filter.
Thus, we used a two-dimensional Butterworth filter in the spatial frequency dimension to suppress  the high-frequency noise present in the data cube, preserving the frequencies containing the astrophysical information. The details on the construction of the Butterworth filter can be found in Sect.\,5 from \citet{Menezes14}.
We used a Butterworth filter with order $n$\,=\,2 and cut-off frequencies defined in terms of the Nyquist frequency of the data ($R_{\rm Ny}$), corresponding to the highest-order sampled by the Fast Fourier Transform (FFT).
Due to the different size of the original spaxels in the $x$- and $y$-direction of the NIFS data (0\farcs103 and 0\farcs043, respectively), the cut-off frequencies must be slightly different in each direction.
The determination of the cut-off frequency is a crucial step and must be performed carefully: if overestimated, it partially corrects the high-frequency noise that should be suppressed; if underestimated, it filters frequencies containing the astrophysical signal, causing an enlargement of the Point Spread Function (PSF).

\begin{figure}
	\includegraphics[width=\columnwidth]{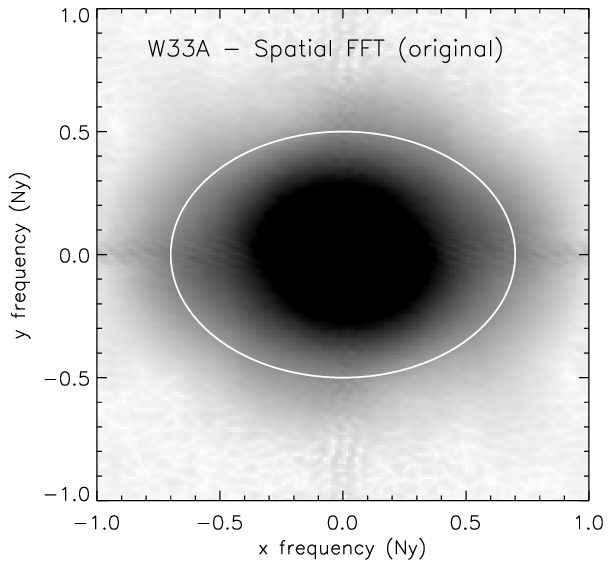} \\[-7.5ex]
	\includegraphics[width=\columnwidth]{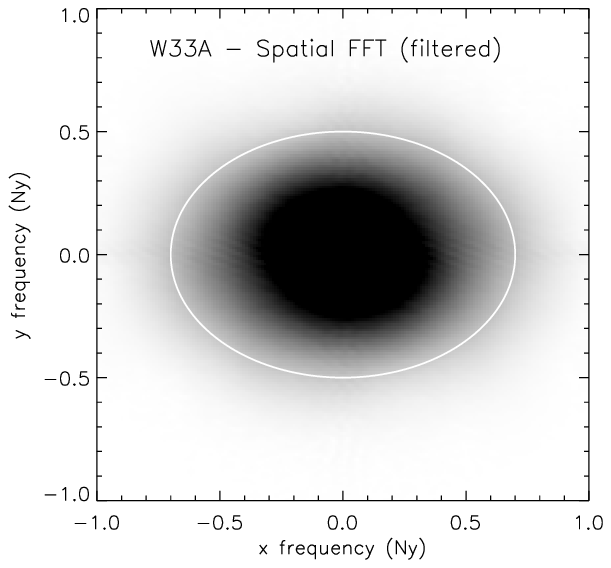} \\[-5.5ex]
    \caption{Mean Fourier Transform of the spatial dimensions of the data cube before (top panel) and after the Butterworth filtering (bottom). The white ellipsoid indicates the cut-off frequency used for constructing the Butterworth filter to suppress the high-frequency noise of the data.}
    \label{fig_spatial_butterworth}
\end{figure}

For W33A observations, frequencies greater than $f_{\rm cutoff,x}$\,=\,0.5\,$R_{\rm Ny}$ and $f_{\rm cutoff,y}$\,=\,0.7\,$R_{\rm Ny}$ were filtered.
Figure\,\ref{fig_spatial_butterworth} presents the mean spatial FFT of the W33A data cube before and after the Butterworth filtering. The cut-off frequencies define an ellipsoid in the Fourier space, indicated by the white contour on top of the mean FFT maps.
We note that almost all the high-frequency features shown in the top panel of Fig.\,\ref{fig_spatial_butterworth} were suppressed by the Butterworth filtering as indicated in the bottom panel of the figure.

\begin{figure}
	\includegraphics[width=\columnwidth]{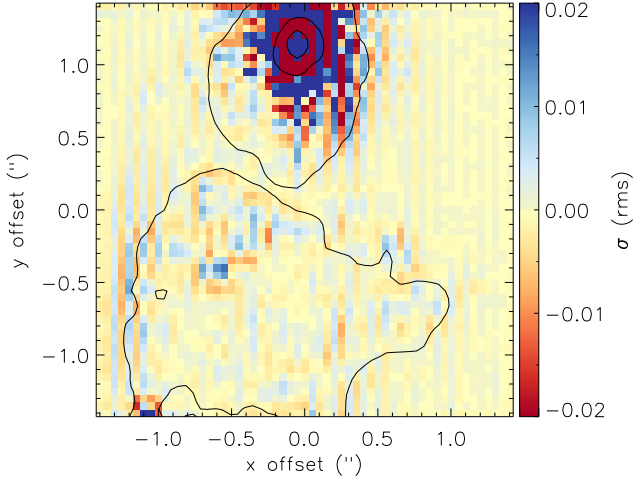} \\[-4.5ex]
    \caption{Difference between the NIFS data cube before and after the spatial Butterworth filtering. The residuals are shown in units of $\sigma$, in a red-to-blue colour-scale to highlight the details of the noise structure. The contours corresponds to the 1, 10 and 50\% of the peak flux in the original data cube.}
    \label{fig_spatial_butterworth_residuals}
\end{figure}

A simple subtraction of the original cube and the filtered cube can be used to check if the Butterworth filtering was successfully applied. 
Figure\,\ref{fig_spatial_butterworth_residuals} exhibits the residuals between the average of the filtered and the original data cubes.
Since the noise is proportional to the signal-to-noise ratio of the data, the residuals are larger towards the point-like source in the FOV.
Despite this, the residuals exhibit a noisy pattern over the entire FOV.
Even though most of the residuals correspond to only a fraction of the noise of the datacube, the noise suppression makes a significant difference for the analysis of weak spectral features (e.g. the CO absorption features at 2.3\,{\micron}).

\subsection{The PCA Tomography}
\label{sec_pca}

This Section presents a brief description of the Principal Component Analysis (PCA) Tomography\footnote{the routines can be downloaded in \url{http://www.astro.iag.usp.br/~pcatomography/}} technique and its interpretation. Further details can be found in \citep{Steiner09} and references therein.

The PCA is a statistical procedure that transforms a set of multi-dimensional variables which are mutually dependent (i.e. correlated) into a set of orthogonal and linearly uncorrelated variables.
The PCA Tomography is applied to the data cubes in order to quantify and extract information, ordered by the fraction of the variance of these new variables: the first of them explains the highest amount of the data variance, and the higher-order variables explain successively smaller amounts of the data variance.
In the transformed coordinate system, the spectral pixels of the data cube correspond to the variables, and the spaxels are the observables.
The new set of uncorrelated variables -- the eigenspectra -- are obtained as a function of the wavelength. The projection of the observables (that is, the spaxels) on the eigenspectra are images, referred as tomograms.

The new coordinates of the data are called Principal Components (PCs), and each PC is orthogonal (i.e. uncorrelated) to the others. Each principal component (PC) is composed of:
$i)$ an eigenvalue, $A_i$, which indicates the relevance of the PC in explaining the information contained in the original data cube; 
$ii)$ a tomogram, $T_i(x,y)$, corresponding to the projection of the PC on the spatial dimensions of the image; and
$iii)$ an eigenspectrum, $E_i(\lambda)$, which is the projection of the PC on the spectral direction.
These three components must be analysed together to properly evaluate the information contained in each PC.
The maximum number of PCs corresponds to the number of the spectral elements of a given data cube. 

The PCA also works as a linear dimensional reduction technique by compressing the data, expressed initially as a large set of correlated variables, into a small but optimal set of uncorrelated variables, ordered by their eigenvalues.
Hence, most of the information of the original dataset can be explained with a relatively small number of PCs.
The analysis of the relative value of $A_i$ as a function of the PC number -- the so-called Scree test -- allows us to determine the best number of PCs that explains the initial data (see Sect.\,\ref{sec_scree}).

Due to the orthogonality of the PCs, the original data cube can be reconstructed by considering a smaller set of PCs with insignificant losses of its astrophysical information. This procedure allows us to suppress a significant amount of noise which is often associated with high-order PCs (see Sect.\,\ref{sec_reconstruction}).

\subsubsection{The Eigenvalues and the Scree Test}
\label{sec_scree}

Each Principal Component is ordered by its eigenvalue (largest to smallest), $A_i$, which indicates the relevance of the PC in explain the information contained in the data cube.
In other words, the eigenvalue is a direct measure of the variance contained in the PC.
The fractional variance of each principal component $i$, $V_i$, is expressed as the ratio between its eigenvalue $A_i$ and the sum of the eigenvalues of all the PCs, given by:
    \begin{equation}
         V_i = \frac{ A_i }{\displaystyle\sum\limits_{i=1}^N A_i }
         \label{eq_variance}
    \end{equation}
    
Table\,\ref{tab:scree_science} lists the fractional and cumulative variance for selected Principal Components of the PCA Tomography analysis of W33A. We note that the first PC contains about 99.8\% of the total variance of the data cube, while higher-order PCs exhibit successively smaller fractional variance values. For W33A, the maximum number of components is equal to 1782, which is the number of spectral pixels of the data cube. 

\begin{table}
    \centering
    \caption{Scree test of the PCA tomography for the \K-band data cube of W33A.}
    \label{tab:scree_science}
    \begin{tabular}{r|rr}
    \hline
\multicolumn{1}{c}{N$_i$}   &   \multicolumn{1}{c}{Variance} & \multicolumn{1}{c}{Cumulative}    \\
        &    \multicolumn{1}{c}{(\%)}    & \multicolumn{1}{c}{Variance (\%)} \\
    \hline
      1 &   99.80360 &   99.8036    \\
      2 &    0.18817 &   99.9918    \\
      3 &    0.00253 &   99.9943   \\
      4 &    0.00109 &   99.9954    \\
      5 &    0.00042 &   99.9958    \\
      6 &    0.00025 &   99.9960    \\
      7 &    0.00021 &   99.9962    \\
      8 &    0.00013 &   99.9964    \\
      9 &    0.00008 &   99.9965    \\
      \ldots & &  \\
      260 &   5.06$\cdot$10$^{-6}$ &   99.999 \\
      \ldots & &  \\
      1782 &   4.61$\cdot$10$^{-10}$ &  100.00 \\
    \hline
    \end{tabular}
\end{table}     

\begin{figure}
    \includegraphics[width=\columnwidth]{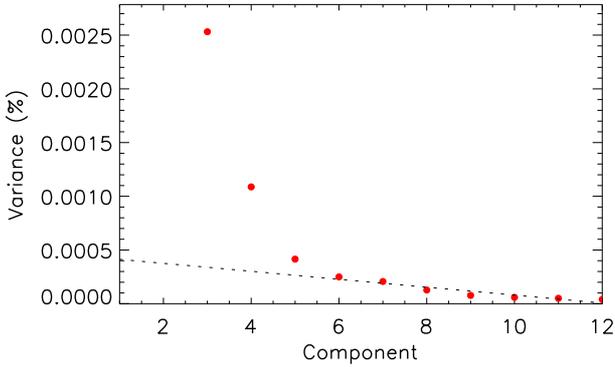} \\[-2.5ex]
    \caption{Scree test showing the variance of each Principal Component (PC) versus the Principal Component order. The variance of each PC is presented as a percentile of the total variance of the data cube (see Eq.\,\eqref{eq_variance}). A dashed line indicates the noise level, based on the linear fit of the data associated with PCs of order 10 or higher.}
    \label{fig_screetest}
\end{figure}

The analysis of the fractional variance alone does not provide a useful way to determine how many PCs contains astrophysical information in the datacube. 
This diagnosis can be assessed by constructing the Scree test of the PCA, shown in Fig.\,\ref{fig_screetest}, which presents the distribution of the fractional variance as a function of the order of the PC.
The Scree test shows a rapid decrease of the variance as a function of the number of the PC, exhibiting two distinct regions: the \textit{signal-dominated} PCs (from PC1 to PC5) from which the variances are significantly larger than the noise level, and the \textit{noise-dominated} PCs (PC6 and onward).
The noise level was estimated by performing a linear fit of $V_i$ by $i$ for the components dominated by noise (PC\,$\geq$\,10).

The analysis and interpretation of the tomograms and eigenspectra of the Principal components  are presented in Sect.\,\ref{sec_pcaresults}.

\subsubsection{Noise suppression and reconstruction of the data cube}
\label{sec_reconstruction}

The first 5 PCs contain about $\approx$99.996\% of the input variance of the data cube (Table\,\ref{tab:scree_science}) and the individual variance of these PCs are above the noise level (see Fig.\,\ref{fig_screetest}). Thus, we reconstructed the datacube using the first five PCs, which led to the suppression of a small but important amount of noise from the initial data cube.

\begin{figure*}
	\includegraphics[width=0.85\linewidth]{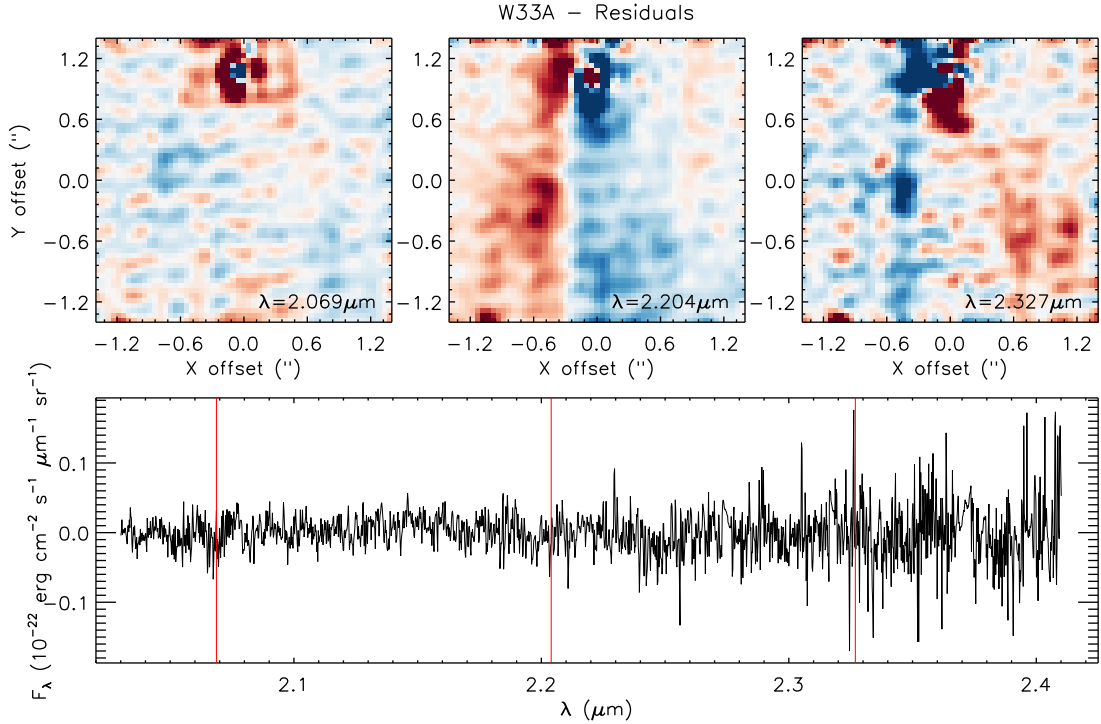} \\[-3.0ex]
    \caption{Residuals between the original W33A data cube and the reconstructed data cube using the first 5 PCs.
    The top panels present the spatial distribution of the residuals of individual spectral channels indicated by the vertical filled red lines of the mean spectrum shown on the bottom panel.}
    \label{fig_W33A_residuals}
\end{figure*}

Figure\,\ref{fig_W33A_residuals} exhibits the spatial and spectral residuals between the original and the reconstructed data cube.
Cuts of the residuals in the spatial and spectral dimension exhibit a large-scale noise pattern distributed over the whole FOV, with relatively stronger intensity closer to the bright point-like source.
These noise patterns are often referred as ``instrumental fingerprints" \citep[see details in Sect.\,6 from][]{Menezes14} and they correspond to a systematic noise structure within IFU data cubes.
As expected, despite the presence of the instrumental fingerprints in the residuals, no coherent spatial or spectral structures related to the astrophysical object were identified.

\section{Results}
\label{sec_results}

\subsection{The large-scale near-infrared view of W33A}
\label{sec_jhk}

The analysis of the large-scale \K-band emission of W33A was presented by \citet[][{hereafter} \Davies]{Davies10}, exhibiting the embedded protostellar object W33A and its extended \K-band nebulosity (see their Fig.\,1).
We took advantage of the available $JHK$ and \hh maps of the UWISH2 survey \citep{Froebrich11} to present a more complete view of the infrared content associated with W33A, including any large-scale \hh emission at 2.12\,{\micron}, a clear signature of star-forming activity for a broad range of bolometric luminosity values \citep[e.g.][]{Varricatt10,Navarete15}.

\begin{figure}
	\includegraphics[width=\linewidth]{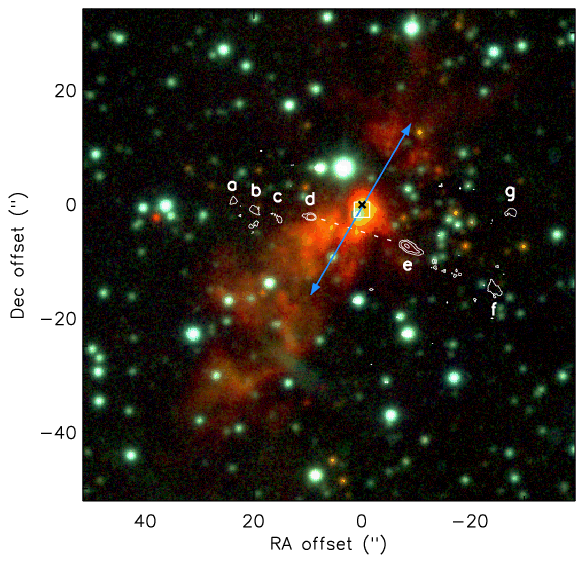}   \\[-5.0ex]
	\includegraphics[width=\linewidth]{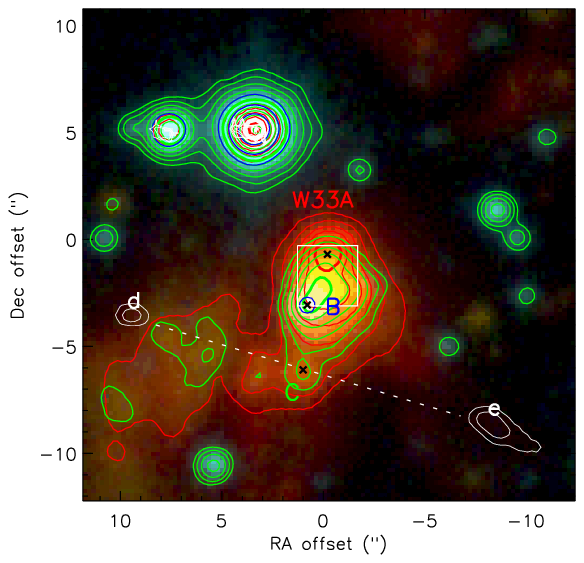}  \\[-7.0ex]
    \caption{Top panel: False-colour $JHK$ map (blue: $J$, green: $H$, red: \K) around W33A from the UWISH2 data. The position of the NIFS map is indicated by the white box. W33A is indicated by the black $\times$ symbol{, corresponding to the (0,0) position in the map (RA\,=\,18:14:39.53, Decl.\,=\,$-$17:52:00.0)}. The rotation axis of the disc reported by \citet{Davies10} is indicated by the blue arrow. The white contours indicate the emission of the continuum-subtracted \hh image at the levels $\sigma^n$ (where $n$\,=\,$1,2,3...$). Knots of \hh emission are labelled in white, and the white dashed line connects the knots labelled as ``d'' and ``e''.
    Bottom panel: A closer view of the $JHK$ map in a 24\arcsec$\times$24\arcsec box around the W33A. The RGB map is overlaid by contours of each near-infrared filter (blue: $J$, green: $H$, red: \K) to highlight the details closer to the brightest region of the main nebula. The \hh knots ``d'' and ``e'' are indicated by the white contours, together with a dashed white line connecting them. Additional point-like sources are indicated by black $\times$ symbols and labelled as B and C, respectively.}
    \label{fig_nir_w33a}
\end{figure}

Figure\,\ref{fig_nir_w33a} presents a false-colour $JHK$ map of a 1.5\,arcmin$^2$ region around W33A from the UWISH2 data, overlaid by contours of the continuum-subtracted \hh emission. 
The top panel, shows the \K-band reflection nebula {illuminated} by W33A, oriented in the SE-NW direction.
The weakness of the NW lobe compared to the SE is then due to the orientation of the system combined with a strong extinction gradient from the (SE) blue- to the (NW) red-shifted lobe of the \K-band outflow (in Sect.\,\ref{sec_h2jet}, we show that the \hh emission identified is blue-shifted in the NIFS FOV, and the presumptive red-shifted lobe is off the NIFS FOV to the NW).
The NIFS FOV is indicated by the white box, the position of the \K-band compact source is indicated by the black $\times$ symbol, and the position angle ({PA\,=\,150$^\circ$}) of the disc reported by \Davies is shown by the blue arrows, which is aligned with the \K-band nebulosity.

The white contours on top of the NIR map indicate the continuum-subtracted \hh emission. The \hh contours indicate no clear extended emission aligned with the main \K-band nebulosity of W33A.
Instead, they are tracing a set of \hh knots roughly aligned in the E-W direction, as previously identified by \citet{Lee13} {(see their Fig.\,2)}.
At least six bright \hh knots (labelled from ``a'' to ``f'') were identified as arising from a bipolar jet. 
The jet has a projected length of {$\sim$50\arcsec}, corresponding to a linear size of 0.58\,parsecs at the distance of W33A (2.4\,kpc).
The large-scale \hh jet is offset by a few arcseconds to the S direction of W33A, suggesting that at least another active protostar is located in the vicinity of W33A.

To better constrain the origin of the \hh jet, a zoomed in view of a {24\arcsec\,$\times$\,24\arcsec} region around W33A is shown in the bottom panel of Fig.\,\ref{fig_nir_w33a}.
Contours of the $J$-, $H$-, and $K$-band filters (blue, green, and red, respectively) are overlaid to highlight the structures around the W33A source, including two additional point-like sources located a few arcseconds to the S of W33A (labelled as B and C), and an extended $H$-band emission oriented to the NW direction.
Source\,B (RA\,=\,18:14:39.584, Dec.\,=\,$-$17:52:02.71) 
is likely driving an extended $H$-band emission oriented to the NW direction, overlapping the $K$-band nebular emission arising from W33A.
Source\,C (RA\,=\,18:14:39.606, Dec.\,=\,$-$17:52:06.06) is observed as a relatively weak compact emission in the $H$-band, located at the central position of the line connecting the \hh knots ``d'' and ``e'', suggesting this object is actually launching the large-scale \hh knots indicated by the white contours.
Despite the fact that Sources\,B and C do not exhibit unequivocal \K-band counterparts in the large-scale maps, any \K-band emission associated with Source\,B and its nebulosity might still be detected in the NIFS FOV.

\begin{figure}
	\includegraphics[width=\linewidth]{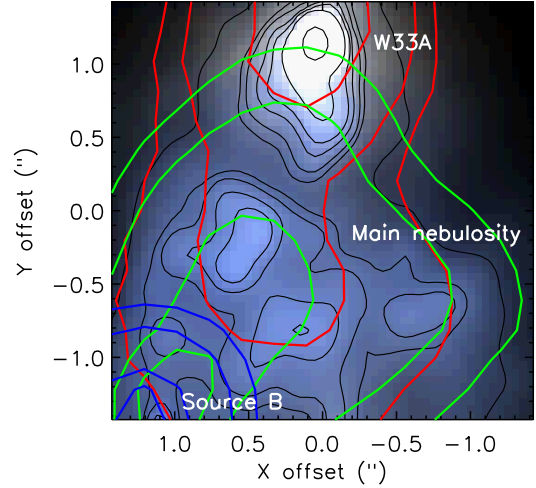}   \\[-5.0ex]
    \caption{False colour composite map of the NIFS field-of-view (blue: 2.05\,{\micron}, green: 2.20\,{\micron}, red: 2.35\,{\micron}) overlaid by the $JHK$ contours (in blue, green and red, respectively). The $JHK$ contours are placed at 25, 40, {75, and 90\%} of the peak intensity of each map. For comparison, the black contours delineate the \K-band emission of the NIFS observations. The main structures are labelled on the image.}
    \label{fig_nifs_jhk}
\end{figure}

In Fig.\,\ref{fig_nifs_jhk} we present the NIFS FOV overlaid by contours of the $JHK$ filters. Thanks to the relatively small plate scale of the {NIR} maps (0\farcs2\,pixel$^{-1}$), we were able to extract maps with the same size as the NIFS FOV, allowing us to compare the high-resolution NIFS data (0\farcs15) with coarser resolution data taken at shorter wavelengths.
The map exhibits the point-source of W33A at the top central region and the main \K-band nebulosity to the South.
The comparison between the black NIFS \K-band contours and the contours of the $H$- and $K$-band (green and red, respectively) suggests that the main nebulosity may be a superposition of distinct structures within the NIFS FOV: it is partially arising from W33A at the upper region of the FOV, while the \K-band nebula is better correlated with the extended $H$-band emission (green contours) towards the South.

The relatively blue colour of the main nebulosity when compared to W33A also favours that the extended emission is brighter at shorter wavelengths, as expected for the extended $H$-band emission detected in the NIFS FOV.
%
At the bottom left (SE direction from W33A), a faint \K-band point-like source peaks at the {brightest and central region of the $H$-band contours}, matching the expected \K-band counterpart of Source\,B identified in the $JHK$ map (see bottom panel of Fig.\,\ref{fig_nir_w33a}).

A detailed inspection in the vicinity of W33A confirms that multiple (proto)stars are located at the line-of-sight of the main \K-band nebulosity of W33A.
The complementary information from the $JHK$ maps indicates that both the compact emission arising from Source\,B and its extended $H$-band emission {might contribute} to the {extended} structures identified in the NIFS FOV{, as we cannot be sure whether such emission comes from the scattered light in the outflow cavity or from source B.
In addition, the $J-$band counterpart seems to be a chance alignment of an optically visible star as indicated by DSS images (not shown in the present work).}

\subsection{The NIFS K-band spectra of the bright compact source and the nebular emission in W33A}
\label{sec_nifs_analysis}

In this section, we briefly analyse the overall properties of the \K-band NIFS observations of W33A to guide the reader through the PCA Tomography analysis presented in Sect.\,\ref{sec_pcaresults}.

\begin{figure*}
	\includegraphics[width=\linewidth]{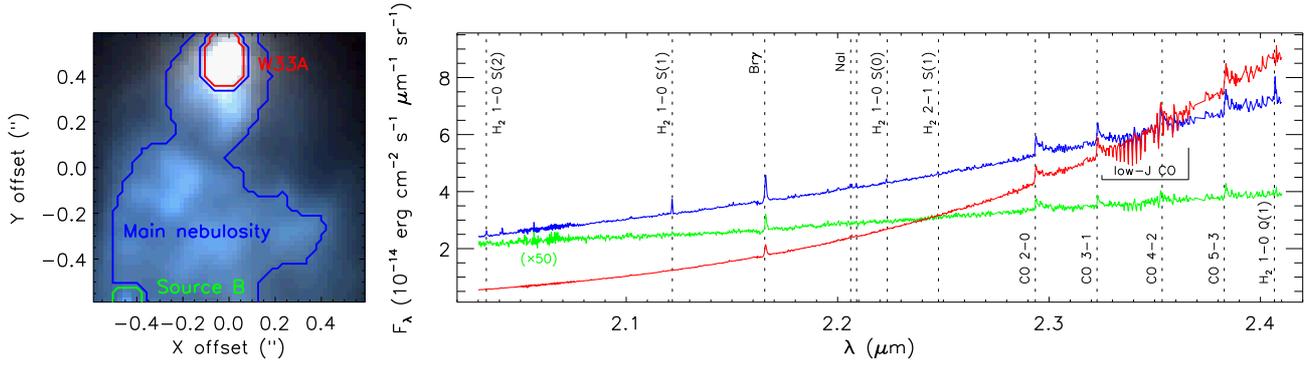}\\[-2.5ex]
    \caption{Left panel: False-colour RGB image of the NIFS FOV (left, blue: 2.05{\micron}, green: 2.12{\micron}, red:2.35\micron) using the flux-calibrated data cube.
    The contours indicate the regions where the integrated {\K-band spectra (right panel) were extracted: the W33A protostar (in red), the main nebulosity (in blue) and the compact emission corresponding to Source\,B (in green).}.
    Right: {Integrated \K-band spectra of each region indicated in the FOV. The spectrum of Source\,B (in green) was multiplied by a factor of 50. No nebular contribution was subtracted.}
    The main spectral transitions present in the data are indicated by the vertical dashed lines: the \hh transitions, the \brg feature at 2.16\,{\micron}, the \nai doublet at 2.21\,{\micron}, the CO bandhead features in emission at $\lambda$\,$\>$\,2.29\,{\micron}, and the CO absorption features at $\lambda$\,$\sim$\,2.32\,{\micron}.}
    \label{fig_W33A_nifs_original}
\end{figure*}

The left panel of Figure\,\ref{fig_W33A_nifs_original} presents a composite RGB map of the NIFS FOV based on the integrated emission of a 0.4\,{\micron} band-pass centred at 2.05 (blue), 2.12 (green), and 2.35\,{\micron} (red).
The red and blue contours correspond to the regions used for integrating the spectrum of the W33A protostar and the extended nebulosity, respectively, shown in the right panel.
The spectra shown in the right panel were integrated over the regions indicated by the contours: the W33A protostar (red), the main nebulosity (blue) and the spaxels corresponding to Source\,B (in green).
We excluded the contribution of Source\,B (the corresponding spaxels are flagged by the green contour in the left panel) from the nebular emission. The \K-band spectrum of Source\,B (shown as green) is about $\sim$100 times fainter than the extended nebula (in blue), and does not significantly affect the integrated spectrum of the brighter region. The spectrum of Source\,B will be further discussed in Sect.\,\ref{sec_nifs_sourceB}.

The integrated spectra indicate that the compact source shows a steeper increase in flux towards larger wavelengths when compared to the nebulosity, as expected for enshrouded protostellar objects. In addition, the nebulosity is brighter than the protostar at short wavelengths, suggesting a relatively strong local reddening effect by the dust towards W33A, that decreases at larger distances from the source.
This may be due to the viewing geometry. Since the compact source is likely seen through the obscuring disc/torus, it appears more reddened than the cavity of the outflow (i.e. the nebulosity) we see in scattered light. The low-$J$ CO absorption lines support this picture since they are strong toward the point source and missing from the nebular spectrum as expected if dense cool gas surrounds the point source.

Both the protostar and the nebulosity shows the presence of the \brg feature in emission at 2.166\,{\micron}, the first four CO bandhead features in emission at $\lambda$\,$\>$\,2.29\,{\micron}, and {the R- and P-branches of the} CO absorption features between 2.32 and 2.37\,{\micron}. A weak emission of the \nai doubled at 2.21\,{\micron} is also observed in both spectra.
At least five ro-vibrational transitions of the \hh molecule are  only detected towards the extended nebula, which is also roughly oriented in the direction of an extended \hh {emission driven} by W33A as reported by \Davies.

\subsection{The PCA Tomography of the \texorpdfstring{$K$}{K}-band NIFS data cube} 
\label{sec_pcaresults}

In this section, we present the analysis and interpretation of the first five Principal Components of the PCA tomography applied to the \K-band data cube of W33A.
The tomograms and eigenspectra associated to these PCs are shown in Figs.\,\ref{fig_pca_final_PC1} to \ref{fig_pca_final_PC5}.

\subsubsection*{PC1}

In general, the overall spatial and spectral information from PC1 roughly correspond to projections of the integrated properties of the data cube {(i.e. its mean in the spatial and spectral dimensions). However, these projections are not given in positive or negative fluxes, but rather in correlations and anti-correlations.
Such a strong resemblance to the data is expected since} PC1 contains the largest relative variance of the data cube (explaining roughly 99.8\% of the data; see Table\,\ref{tab:scree_science}.

\begin{figure*}
\centering
    \includegraphics[width=\linewidth]{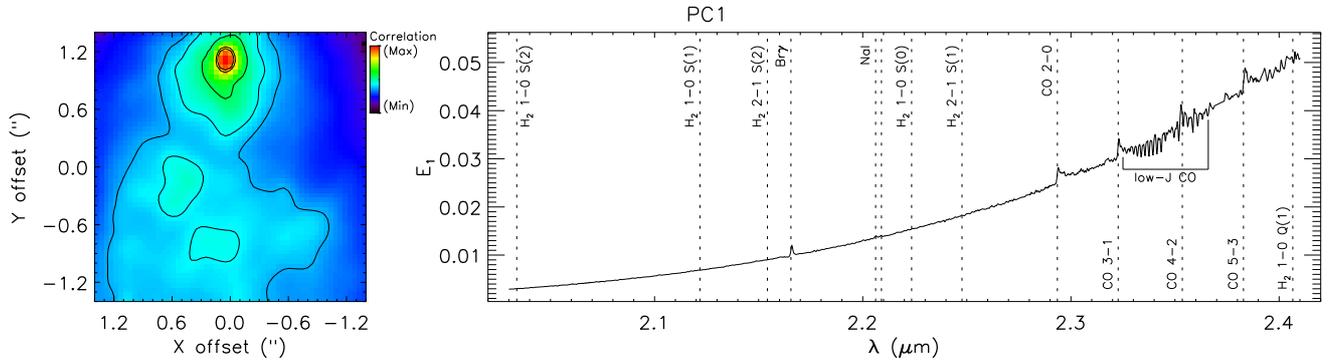} \\ [-2.0ex]
    \caption{The tomogram (left panel) and the corresponding eigenspectrum (right) of the Principal Component 1 of the PCA Tomography of W33A.
    The variance of the tomogram is indicated in a rainbow colour scale, where black indicates weak correlation (min), and red indicates strong correlations (max).
    The contour-levels are chosen to highlight the extended emission and the point-like source. The eigenspectrum exhibits the variance ($E_1$) as a function of wavelength. The position of spectral features guiding the analysis of the data are labelled and indicated by vertical dashed lines.}
    \label{fig_pca_final_PC1}
\end{figure*}

The tomogram of PC1 exhibits a peak at the position of the bright point-like source corresponding to the W33A protostar.
{The main nebulosity is also identified in the tomogram, exhibiting larger correlation at the vicinity of W33A and decreasing towards the South.} 
The correlation between the compact and the extended emission (i.e. the fact that both appear as positive variances in the tomogram) indicates that both structures shares the same spectral information shown in the eigenspectrum, shown in the top right panel of Fig.\,\ref{fig_pca_final_PC1}.

The eigenspectrum shows that the variance of the continuum emission increases towards larger wavelengths. Such behaviour (in the flux spectrum) is often associated with thermal emission of a relatively hot source (dust), as expected for a protostar.
Discrete spectral features such as the \brg, the CO bandhead features, and a relatively weak Na\,{\sc{i}} doublet are also observed with positive variance and, thus, correlated with the continuum emission.
In addition, the low-$J$ CO absorption lines are identified as absorption features.

\subsubsection*{PC2}

The higher-order PCs present the correlation and anti-correlation between the spectral features in the eigenspectra, and their association with the structures probed by the tomograms.
Thus, for better highlighting the variance gradient between distinct structures, the tomograms are shown in a red-to-blue scale, where correlated (positive variance) and anti-correlated regions (negative variance) are shown in blue and red, respectively.

\begin{figure*}
\centering
    \includegraphics[width=\linewidth]{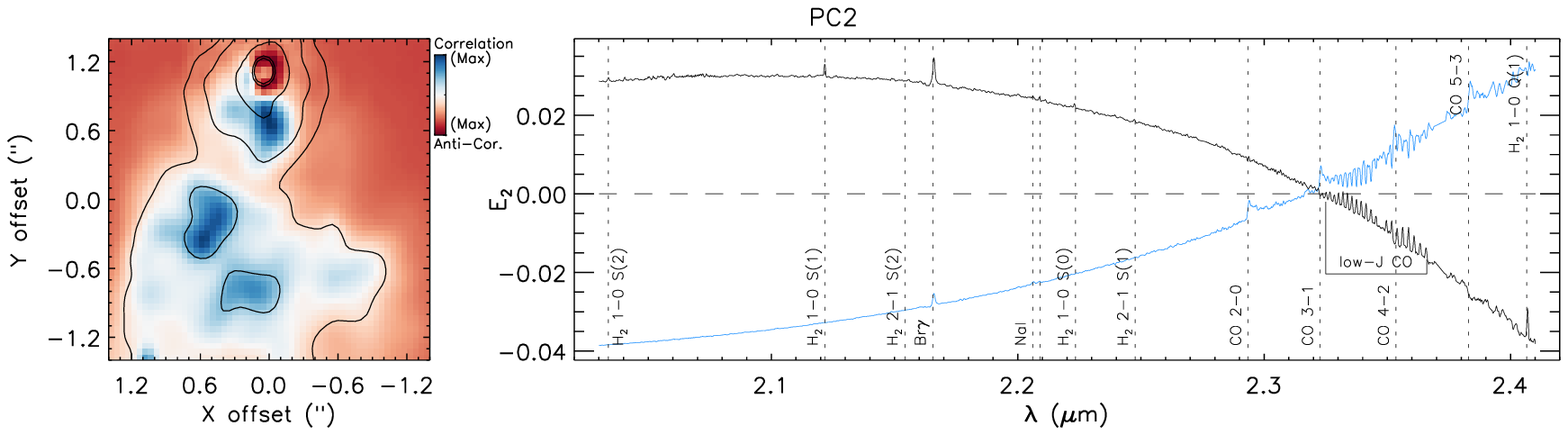} \\
    \includegraphics[width=\textwidth]{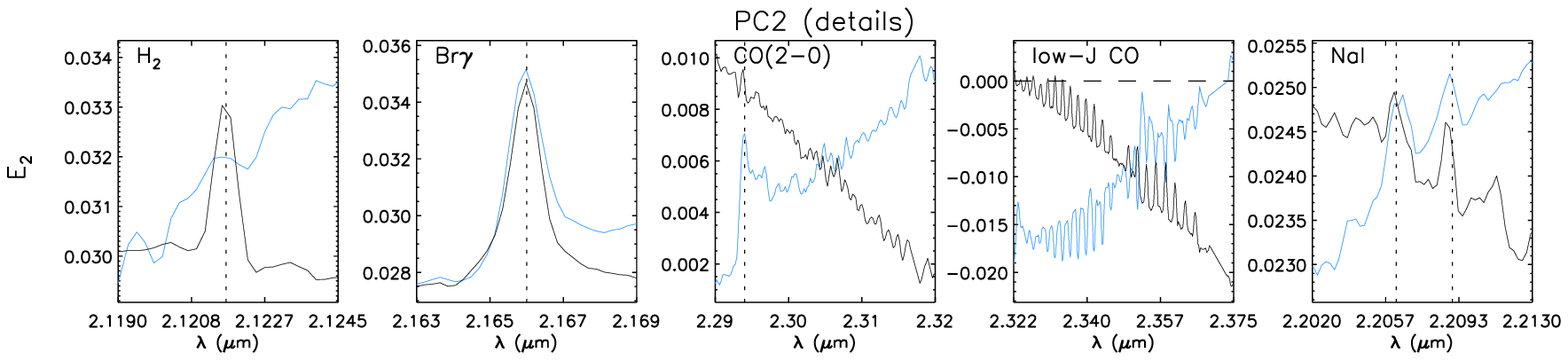} \\[-1.0ex]
    \caption{The tomogram (top left) and the corresponding eigenspectrum (top right) of the Principal Component 2 of the PCA Tomography of W33A.
    In the left panel, the contours of the PC1 tomogram are {overlaid as black curves}.
    For orders $n$\,$\geq$\,2, the tomogram is shown in a red-to-blue colour scale indicated in the top right, where blue indicates the strongest correlations and red indicates the strongest anti-correlations.
    In the right panel, the eigenspectrum associated with PC1 is overlaid as the blue curve. The horizontal dashed line indicates variance equal to zero.
    A detailed view of the variances of selected spectral lines  of the eigenspectrum are shown in the bottom panels (from left to right: the H$_2$ line at 2.12\,{\micron}, the \brg feature at 2.16\,{\micron}, the CO\,(2-0) bandhead emission at 2.29\,{\micron}, the CO absorption features at 2.32\,{\micron}, and the \nai doubled at 2.20\,{\micron}).}
    \label{fig_pca_final_PC2}
\end{figure*}

In the top left panel of Fig.\,\ref{fig_pca_final_PC2}, the tomogram of PC2 exhibits the extended nebulosity with positive variance (shown in blue), anti-correlated with a ring-like structure around W33A exhibiting negative variance values (in red). 
The eigenspectrum shows continuum and discrete spectral features that are related to the structures observed in the tomogram.
The continuum emission shows positive variance values for $\lambda$\,$\lesssim$\,2.32\,{\micron}, and is negative for longer wavelengths. Combining the spectral and spatial information, we interpret the continuum emission as dominated by the main nebulosity at short wavelengths, while the structure around the compact source is brighter at longer wavelengths.
This behaviour can also be interpreted in terms of the local extinction of the dust: the reddening is stronger towards the compact source than that observed in the line-of-sight of the nebulosity. This interpretation is also consistent with the RGB image shown in Fig.\,\ref{fig_nifs_jhk}.

The top right and bottom panels of Fig.\,\,\ref{fig_pca_final_PC2} also exhibits the H$_2$ transitions, \brg and \nai lines correlated with the blue continuum, indicating that these lines are produced in the main nebulosity and absent in the compact source.
\citet{Davies10} presented spectra of the individual bright regions shown in the nebulosity of Figure\,\ref{fig_pca_final_PC2} (see their Figs.\,3 and 10).
The three bright regions in the cavity from PC2 correspond to the regions defined as B, C and D by \Davies: B is closest to the point source and shows weaker low-$J$ CO absorption strength than the point source, and C and D show yet weaker absorption of the narrow CO transitions.
This is consistent with the spots showing a reflected spectrum of the point source.

The {narrow} CO absorption lines detected at $\lambda$\,$\gtrsim$\,2.32\,{\micron} in the eigenspectrum appear as anti-correlated features, just as the continuum emission at longer wavelengths.
The CO absorption features are produced in a relatively dense and cold environment, where the CO molecules are shielded from the radiation of the protostar. These physical parameters are often observed in the outer regions of circumstellar discs, consistent with their association with the anti-correlated structure around the W33A protostar.
{The absence of the CO bandhead features in the eigenspectrum of PC2 is also remarkable. These features are produced at the inner region of the circumstellar disc, at temperatures of around 2,000\,K \citep{Blum04}, unresolved at the spatial resolution of the NIFS observations ($\sim$0\farcs15, or 360\,AU at 2.4\,kpc).
Indeed, the PC2 tomogram exhibits a ring-like structure around the W33A protostar, with variance values dropping to zero towards its central region, which is consistent with the unresolved origin of the CO bandhead emission.}

\subsubsection*{PC3}

The tomogram of PC3 exhibits a spatially resolved structure around the compact source, with coherent changes between positive (blue) and negative variance values (red) {at the central position of the point-like object}. 
In extra-galactic objects, this pattern is often identified in rotating structures, such stellar discs or torus around super-massive black holes \citep[e.g.][]{Ricci14}.
In the context of protostellar objects, it is likely arising from a rotating circumstellar disc, as identified by \citet{Davies10} when analysing the CO absorption features (see their Fig.\,14).
The rotation axis of the structure has a position angle of about $-$45$^\circ$ (from N to E).

\begin{figure*}
\centering
    \includegraphics[width=\linewidth]{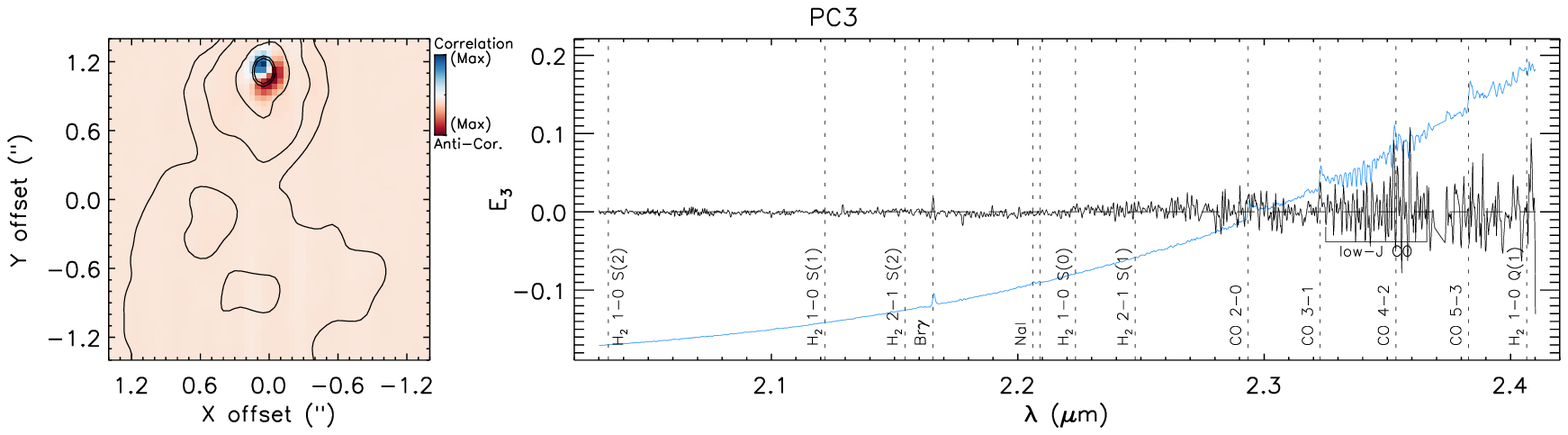} \\
    \includegraphics[width=\linewidth]{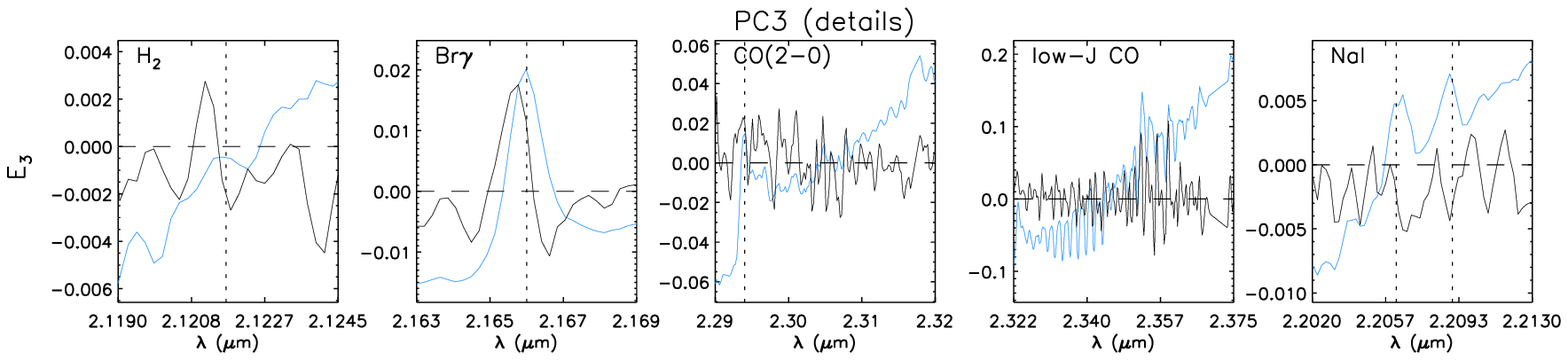} \\[-2.0ex]
    \caption{The tomogram (top left) and the corresponding eigenspectrum (top right) of the Principal Component 3 of the PCA Tomography of W33A. For details, see Figs.\,\ref{fig_pca_final_PC1} and \ref{fig_pca_final_PC2}.}
    \label{fig_pca_final_PC3}
\end{figure*}

The eigenspectrum of PC3 exhibits no variance of the continuum, but shows a series of narrow features for $\lambda$\,$\gtrsim$\,2.3\,{\micron}, likely arising from the low-J CO lines (see details in the bottom panel of Fig.\,\ref{fig_pca_final_PC3}).
The blue-shifted components are correlated and so, associated with the positive variance values in the tomogram (shown as blue) while the red-shifted components are associated with negative values (shown as red), indicating the disc is rotating from NE to SW direction.

A similar kinematic pattern is also observed for the \brg feature (also indicated in the bottom panel of the figure), suggesting that the rotating disc exhibits an ionised component (see discussion in Section\,\ref{sec_discussion_brg}).

\begin{figure}
\centering
    \includegraphics[width=\linewidth]{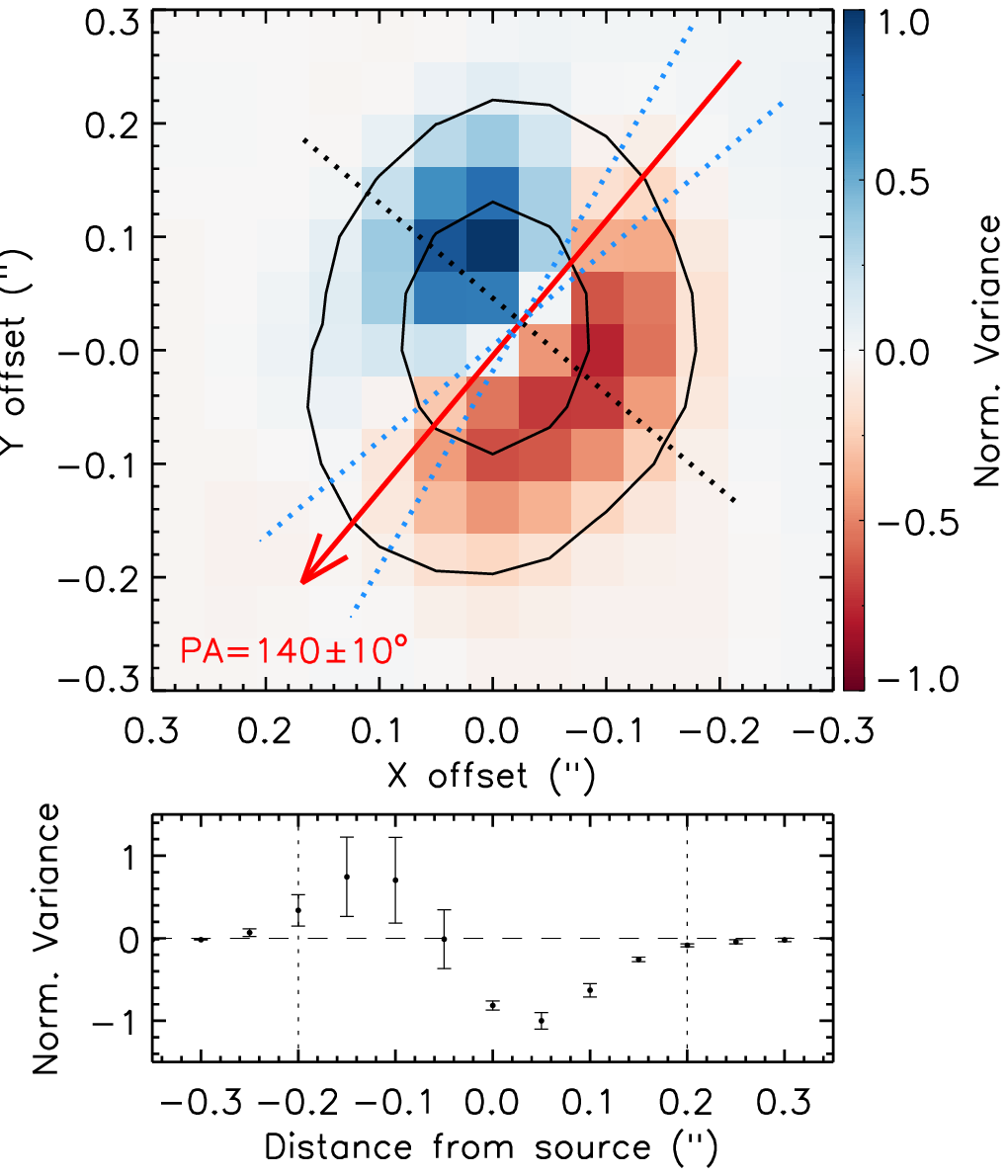} \\[-2.0ex]
    \caption{A detailed view of the disc-like structure probed by the PC3 tomogram. Top panel: The normalised variance of the tomogram is shown in a divergent red-to-blue colour-scale. The position of the compact source is indicated by the black contours. The red arrow and blue dashed lines indicate the {position angle of the} rotation axis of the disc {and its error, respectively, estimated as PA\,=\,140\,$\pm$\,10$^\circ$}. The dashed black line shows the direction from which the radial variance profile was extracted. Bottom: the radial variance profile extracted along the plane of the disc. The points and error bars correspond to the mean variance and the 1-$\sigma$ error of a 3-pixel width region perpendicular to the sampled spatial direction. The dashed vertical lines indicate the region where the absolute value of the variance is larger than zero ($|r|$\,$\leq$\,0\farcs2).}
    \label{fig_PC3_cut}
\end{figure}

Figure\,\ref{fig_PC3_cut} shows a closer view of the rotating structure identified in the tomogram. {The rotation axis of the disc is observed with a position angle of PA\,=\,140\,$\pm$\,10$^\circ$}.
A radial profile of the variance along the disc plane is presented in the bottom panel of the figure, showing that the disc kinematic signature is only observed within {the inner $\pm$0\farcs4 region around W33A (about $\sim$3 times larger than the resolution of the observations)}.
{Even though the CO absorption features are observed in the extended nebulosity, the principal component 3 is only showing the regions exhibiting ordered kinematics on these lines, restricted to the disc. This analysis suggests that the ordered kinematics are observed in a much more compact region than previously reported by \Davies.}

\subsubsection*{PC4} 

The tomogram of PC4 exhibits a correlated structure roughly perpendicular to the rotating disc identified in PC3 (see Fig.\,\ref{fig_pca_final_PC3}).

The eigenspectrum of PC4 shows both continuum and discrete features, as observed in PC2 (Fig.\,\ref{fig_pca_final_PC2}).
The continuum emission between 2.17 and 2.35\,{\micron} is anti-correlated with the emission at shorter and at longer wavelengths.
We interpret this result as if the same region is associated with at least three distinct processes: 
$i)$ the dust content is reflecting the radiation emitted by the protostar at short wavelengths;
$ii)$ the thermal emission of the dust is observed at longer wavelengths;
$iii)$ the same region exhibits the CO bandhead features in emission, and the profiles are quite narrow when compared to the emission observed in PC1. This may indicate that we observe this emission moving perpendicularly to the line-of-sight.

In addition, the eigenspectrum of PC4 shows a strong \brg feature in emission, suggesting the presence of partially-ionised gas that could also contribute as a source of scattering electrons producing the reflected light at shorter wavelengths.

\begin{figure*}
\centering
    \includegraphics[width=\linewidth]{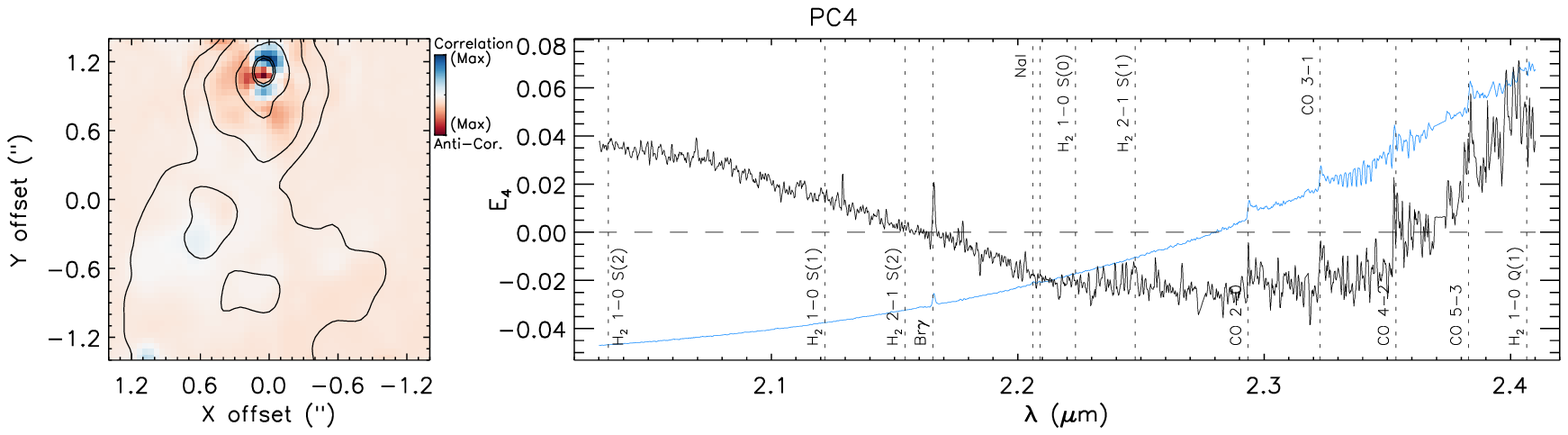} \\
    \includegraphics[width=\textwidth]{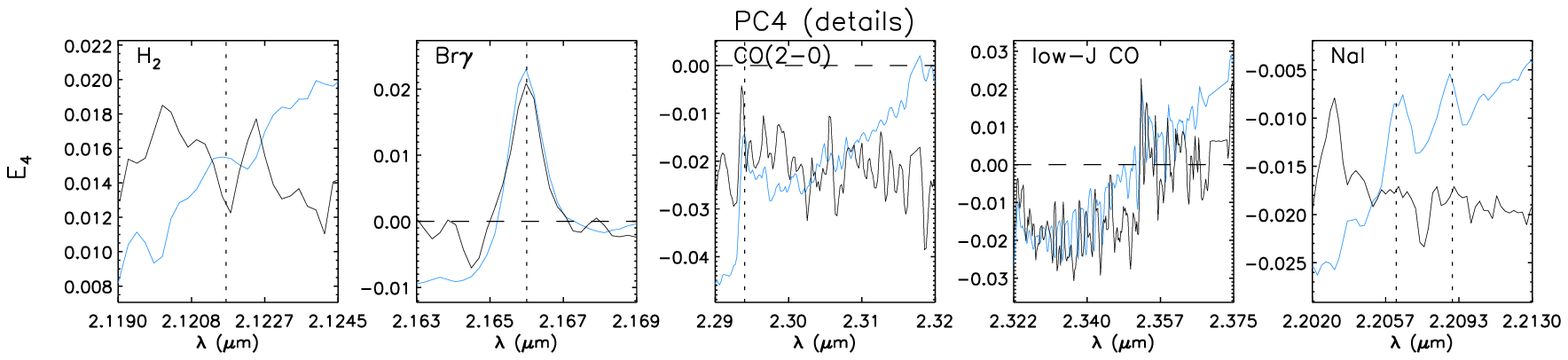} \\[-2.0ex]
    \caption{The tomogram (top left) and the corresponding eigenspectrum (top right) of the Principal Component 4 of the PCA Tomography of W33A. For details, Figs.\,\ref{fig_pca_final_PC1} and \ref{fig_pca_final_PC2}.}
    \label{fig_pca_final_PC4}
\end{figure*}

\subsubsection*{PC5} 

The tomogram of PC5 exhibits an extended emission with a different morphology to the main nebulosity of W33A (shown by the black contours), suggesting that this structure is either produced by a distinct mechanism or exhibits distinct physical conditions. 

Indeed, the eigenvector exhibits at least six \hh transitions correlated with the extended emission shown in the tomogram, indicating that PC5 is {either tracing the base of a disc wind, or the cavity of the large-scale outflow shown in Fig.\,\ref{fig_nifs_jhk},} observed with a position angle (PA) of {145$^\circ$ (from N to E), measured along the brightest and more elongated region of the molecular emission. In addition, the spatial distribution of the \hh emission probed by the tomogram of PC5 is similar to the continuum-subtracted map of the \hh\,(1-0)\,S(1) transition at 2.12\,{\micron} presented by \Davies (see their Fig.\,15). The most significant difference between the tomogram and the line map is that the first considers the emission of all the \hh transitions available in the spectral range (producing a higher signal-to-noise map) while the second method is based only on the intensity of a single line (associated with a relatively larger noise).}

{In addition, the} \brg feature and CO bandhead emission are anti-correlated with the H$_2$ lines, indicating they arise from the compact region at the launching point of the jet {(indicated as red in the tomogram)}.
In general, the CO features are produced in a dense region (e.g. the circumstellar disc) while the \hh is excited through shocks in a more diffuse medium. Therefore, PC5 also shows the spatial separation of regions with distinct density regimes.

\begin{figure*}
\centering
    \includegraphics[width=\linewidth]{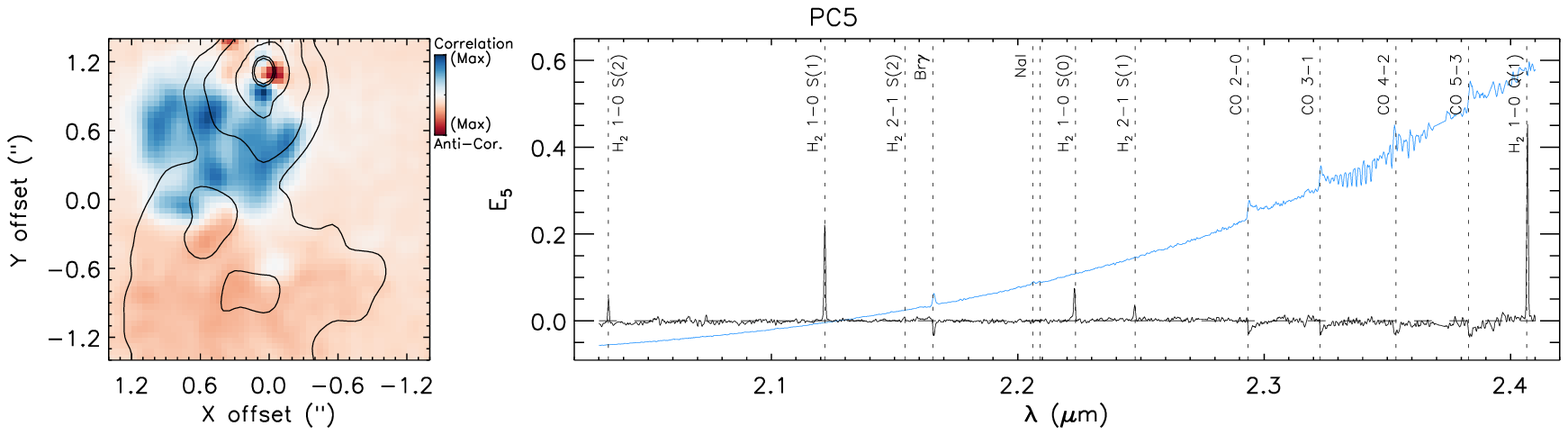} \\
    \includegraphics[width=\textwidth]{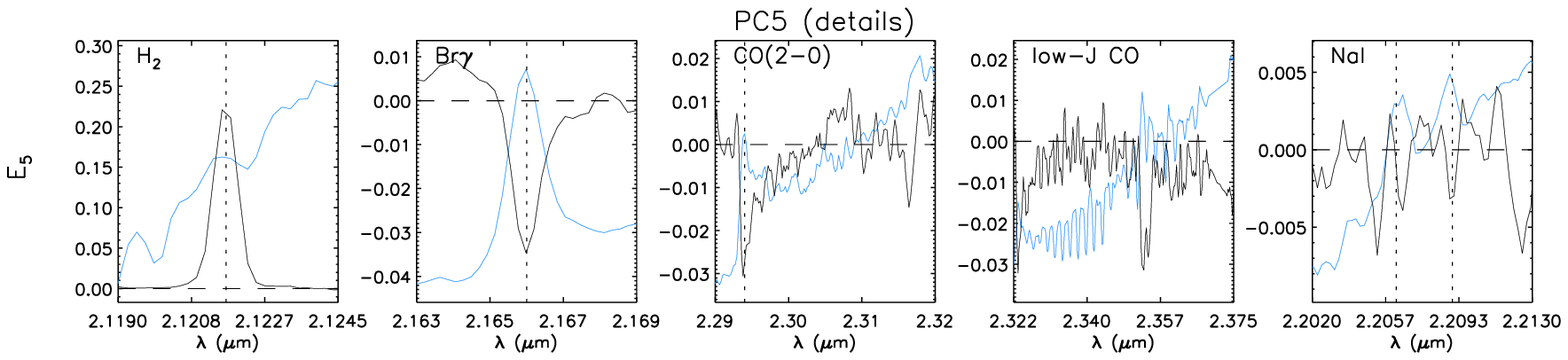} \\[-2.0ex]
    \caption{The tomogram (top left) and the corresponding eigenspectrum (top right) of the Principal Component 5 of the PCA Tomography of W33A. For details, Figs.\,\ref{fig_pca_final_PC1} and \ref{fig_pca_final_PC2}.}
    \label{fig_pca_final_PC5}
\end{figure*}

\subsection{Derivation of the physical parameters of the structures associated with the W33A protostar}
\label{sec_resultsW33A}

The following results were obtained by analysing the data cube reconstructed using the first five Principal Components of the PCA Tomography (Sect.\,\ref{sec_reconstruction}).
We also compare our findings with those reported by \Davies to check whether the post-processing of the NIFS data cubes led to a significant improvement on the determination of the physical parameters.

\begin{figure*}
    \includegraphics[width=\linewidth]{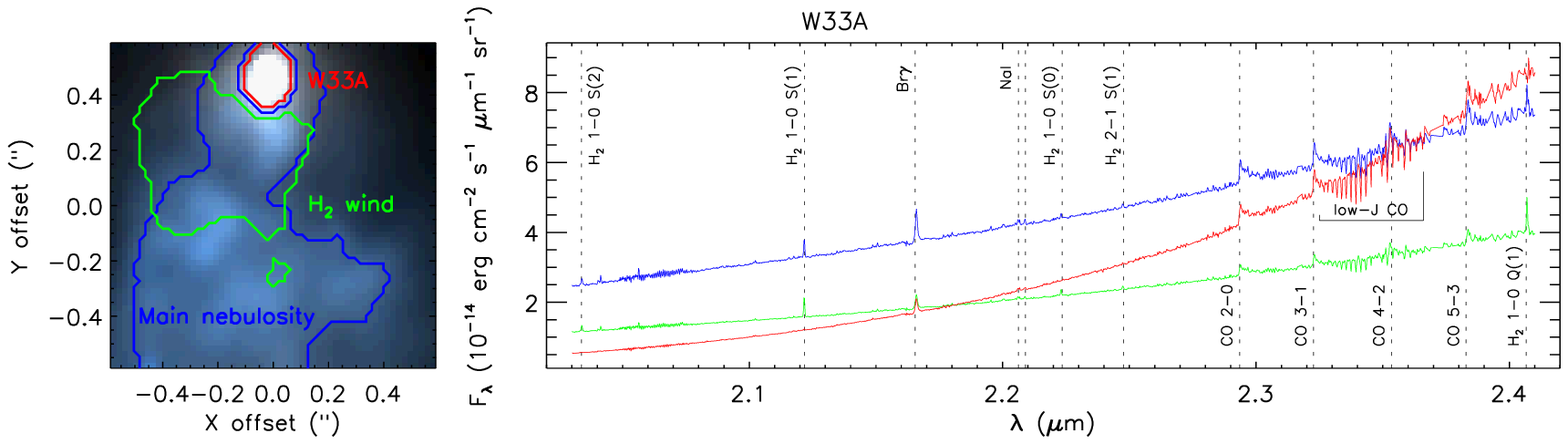}\\[-2.5ex]
    \caption{Left panel: False-colour RGB image of the NIFS FOV (left, blue: 2.05{\micron}, green: 2.2{\micron}, red:2.35\micron) recreated using PCs 1--5.
    The contours indicate the regions where the integrated spectra were extracted.
    Right: integrated \K-band spectra of the W33A point-like source (in red), the main nebulosity (in blue), and the {\hh wind} (in green).}
    \label{fig_W33A_nifs}
\end{figure*}

Figure\,\ref{fig_W33A_nifs} presents a composite RGB image of the NIFS FOV similar to the one presented in Fig.\,\ref{fig_W33A_nifs_original}. 
{The spectra shown in the right panel were integrated over the regions indicated by the contours: the compact \K-band object (red), the {disc wind/cavity of the outflow probed by the molecular \hh emission} (green) and the main nebulosity (blue).}

The integrated spectra shows that the main nebulosity is brighter than the protostar at shorter wavelengths, while the embedded object gets brighter for $\lambda$\,$\gtrsim$\,2.35\,{\micron}.
This behaviour was also observed in the analysis of PC2 (see Fig.\,\ref{fig_pca_final_PC2}), showing that the continuum emission is correlated with the nebulosity at short wavelengths due to the reflection of the radiation from the embedded protostar, while it is correlated with the compact source for longer wavelengths likely due to local extinction effects.

In addition, the spectra of the main nebulosity (in blue) and the {\hh wind} (in green) exhibit at least five ro-vibrational transitions of the \hh molecule. The \hh emission observed in the spectrum of the main nebulosity is likely due to the overlap between this region and the {\hh wind}.
{The most energetic transition corresponds to the H$_2$\,2--1\,S(1) line at 2.2477\,\micron, with upper-level energy of 12,550\,K.}
The absence of transitions with larger energy indicates that the physical conditions of the \hh gas (i.e. temperature and column density) are not sufficient to produce more energetic transitions. We further investigate the physical parameters of the {molecular \hh emission} in Sect.\,\ref{sec_h2jet}.

\subsubsection{The nature of the YSO candidates in the vicinity of W33A}
\label{sec_nifs_sourceB}

{We performed the photometry of the UKIDSS $JHK$ images using the IRAF/DAOPhot package to better characterise the protostellar object candidates in the vicinity of W33A (Sources\,B and C).
We used the {1.4\arcmin\,$\times$1.4\arcmin} FOV shown in the top panel of Fig.\,\ref{fig_nir_w33a}, containing enough point-like sources to obtain a good photometric calibration by cross-matching with the UKIDSS catalogue.}

{First, we obtained an average map of the individual $J$, $H$, and $K$ images to identify the positions of the sources that appears in, at least, one of the three filters. We obtained a list of 298 point-like objects with peak intensities above a 3-$\sigma$ threshold.
We performed the photometry of the individual images, using an aperture with radii of $r$\,=\,{1\arcsec}, and a ring with inner and outer radius of 1\farcs6 and 2\farcs0 to estimate the sky background.
The fluxes were converted to instrumental magnitudes, and the photometric calibration was done by cross-matching the sources detected in all the three bands ($J, H$ and $K$) with the UKIDSS catalogue, leading to a list of 132 objects.
A final list containing the photometry of 287 objects, was obtained after removing the sources detected only in a single filter, excluding, for example, any bright \K-band knot associated with the large-scale outflow in the FOV from the initial list.}

{The $JHK$ magnitudes of the W33A, Source\,B and Source\,C are listed in Table\,\ref{tab_photometry}, and the full list containing the 287 sources is available as on-line material (The first ten rows are listed in Table\,\ref{tab_photometry_all}.
W33A was not detected in the $J$- and $H$-band and, thus, the reported values correspond to upper limits, as well as the $J$-band magnitude of Source\,C. In addition, the reported $J$-band magnitude of Source\,B corresponds to a foreground star in the line-of-sight of the object detected in the $H$- and $K$-bands.}

\begin{table}
    \caption{$JHK$ photometry of the protostellar object candidates in the vicinity of W33A.}
    \label{tab_photometry}
    {\centering
    \begin{tabular}{c|ccc}
    \hline
    \hline
        Source & J (mag) & H (mag) & K (mag) \\
    \hline
      W33A     & 20.77$^\ast$  & 15.05$^\ast$ &  9.51\,$\pm$\,0.19 \\
      Source B & --$^{+}$ & 13.76\,$\pm$\,0.09 & 10.30\,$\pm$\,0.20 \\
      Source C & 20.77$^\ast$  & 16.31\,$\pm$\,0.11 & 12.69\,$\pm$\,0.23 \\
    \hline
    \end{tabular}\\}
    {\textbf{Notes:}
    $\ast$ indicates upper limit due to the non-detection of a point-like source at the given position in the corresponding filter.
    $+$ a foreground source lying very close to the object exhibits $J$\,=\,15.45\,$\pm$\,0.08\,mag.}
\end{table}

\begin{figure}
  \centering
  \includegraphics[width=\linewidth]{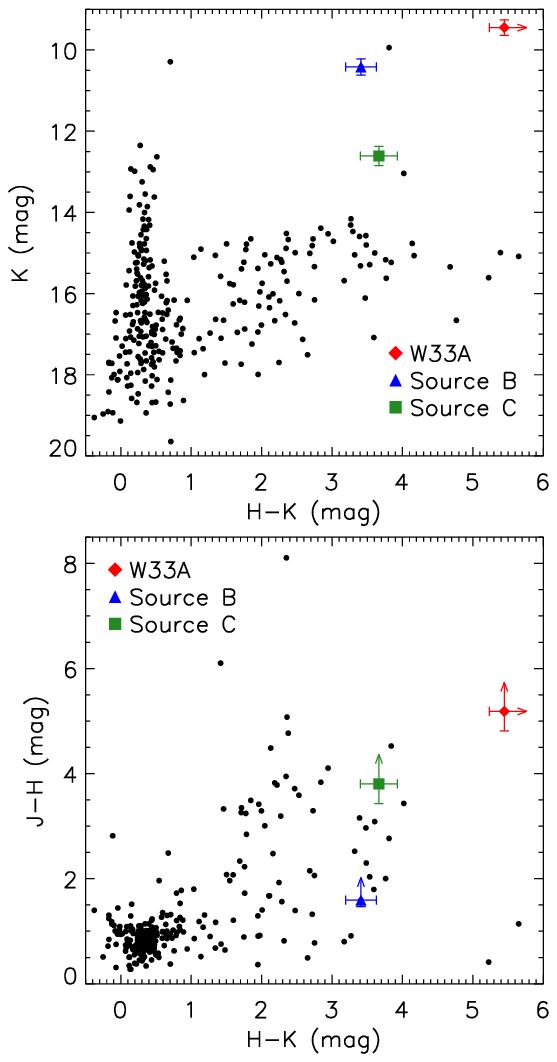}   \\[-2.0ex]
  \caption{$K$ versus $H-K$ colour-magnitude diagram (top panel) and the $J-H$ versus $H-K$ colour-colour diagram (bottom) of the {1.4\arcmin\,$\times$\,1.4\arcmin} region around W33A. W33A is indicated as the red diamond symbol, Source\,B as the blue triangle and Source\,C as the green squared symbol. The photometric errors are shown as error bars and the arrows indicates lower limits of the actual values (see text).}
  \label{fig_cmd_w33a}
\end{figure}

{Fig.\,\ref{fig_cmd_w33a} presents the $K$ versus $H-K$ colour-magnitude diagram (top panel) and the $J-H$ versus $H-K$ diagram (bottom) of the sources within the {1.4\arcmin\,$\times$1.4\arcmin} FOV around W33A. 
The colour-magnitude diagram indicates that the three sources (W33A, Source\,B and Source\,C) are relatively bright \K-band objects ($K$\,$<$\,13\,mag), exhibiting $H-K$ colour indices consistent with embedded young objects ($H-K$\,$>$\,3\,mag).
W33A is the most reddened source, with $H-K$\,=\,5.54\,$\pm$\,0.21\,mag, followed by Source\,C ($H-K$\,=\,3.62\,$\pm$\,0.26\,mag) and Source\,B ($H-K$\,=\,3.46\,$\pm$\,0.22\,mag).
These objects are located among the brightest \K-band sources of a sequence of reddened objects, confirming that they are deeply embedded within their circumstellar environments as expected for protostars.}

The bottom panel of Fig.\,\ref{fig_cmd_w33a} presents a colour-colour diagram based on the $J-H$ and $H-K$ colour indices.
Given that the $J$-band magnitude of the three sources corresponds to upper limits of their real values, their position in the $y$-axis of the diagram correspond to lower limits.
Despite of that, the three sources are consistent with reddened objects, corroborating the young nature of these objects.

\begin{figure*}
	\includegraphics[width=\linewidth]{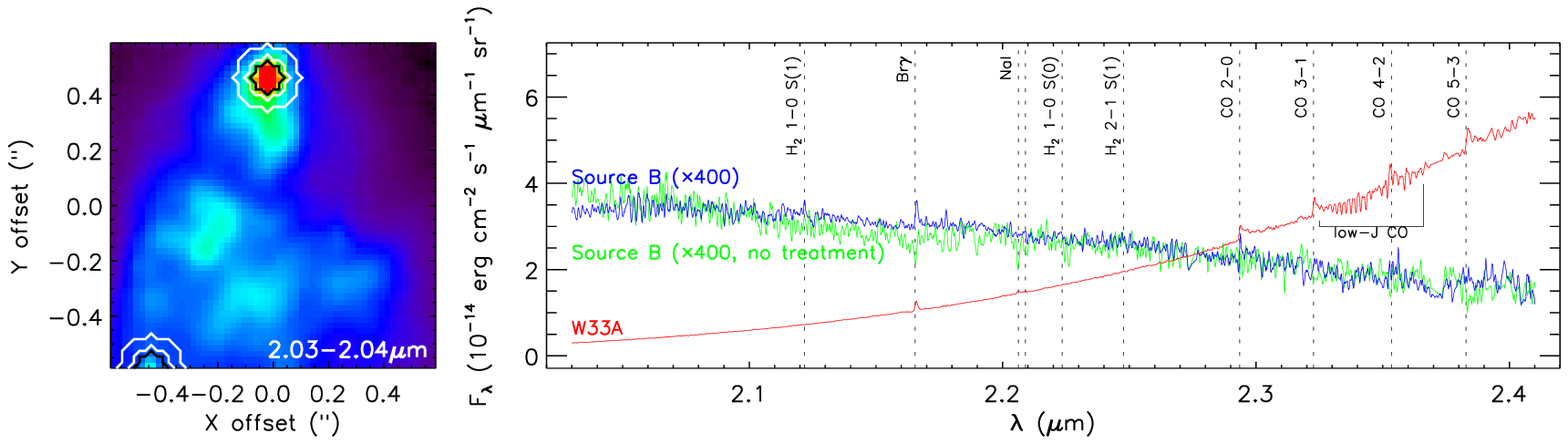}\\[-2.5ex]
    \caption{Left panel: False-colour image of the NIFS FOV integrated from 2.03 to 2.04\,{\micron}. The point-like source is saturated to enhance the contrast between W33A and the fainter, \K-band counterpart of Source\,B at the SE edge of the datacube.
    The black contours indicate the regions where the integrated spectra were extracted. The nebular emission, evaluated over the regions indicated by the white contours, was subtracted from the spectra.
    Right: integrated \K-band spectra of the W33A point-like source (in red), and the Source\,B before (green) and after the post-processing routines applied to the NIFS datacube (in blue).
    The spectra of Source\,B were multiplied by a factor of 400.
    }
    \label{fig_W33A_SourceB}
\end{figure*}

In Fig.\,\ref{fig_W33A_SourceB}, we present the spectrum of W33A (shown as the red curve) and Source\,B (in blue), integrated over a 0\farcs15 radius, each with the extended nebular emission subtracted using an annular region around each compact sources, with inner and outer radii of 0\farcs20 and 0\farcs30, respectively (indicated by the white contours in Fig.\,\ref{fig_W33A_SourceB}).
We note that W33A has a steeper continuum due to the emission of the dust, while Source\,B has a blue continuum.
For comparison, we also show the integrated spectrum of Source\,B extracted using the original data cube (shown as the green curve), before the post-processing of the NIFS data cube and its reconstruction (see details in Sect.\,\ref{sec_postprocess}).
The comparison between the spectrum of Source\,B before and after the reconstruction of the data cube shows the effectiveness of the noise suppression due to the post-processing procedures applied to the data cube. 
In addition, the \K-band spectrum of Source\,B exhibits relatively weak emission of the \brg and the \hh transition at 2.12\,{\micron}. These features are typically observed in YSOs, corroborating the photometric results from Fig.\,\ref{fig_cmd_w33a}.

\subsubsection{The physical parameters of the molecular hydrogen gas}
\label{sec_h2jet}

\begin{figure}
	\includegraphics[width=\linewidth]{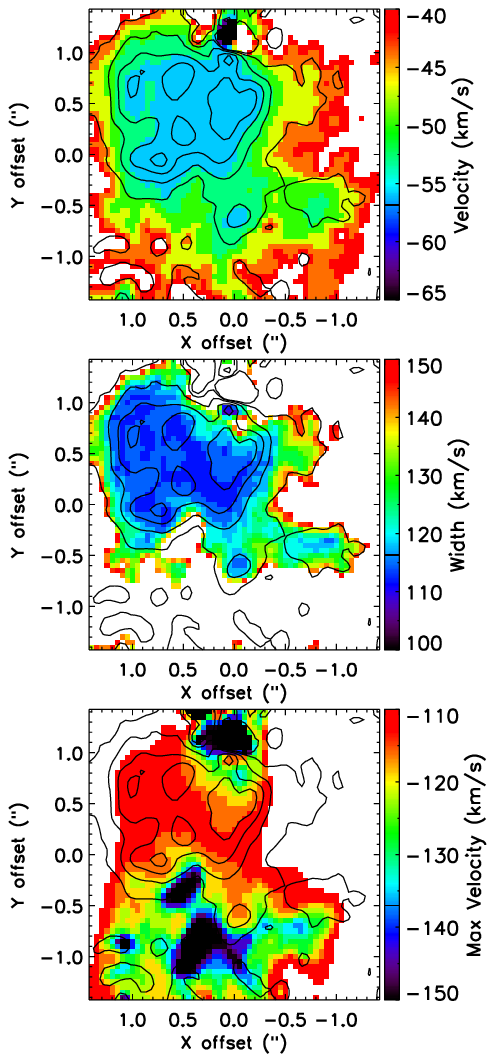} \\[-3.0ex]
    \caption{Velocity map of the \hh\,(1--0)\,S(1) transition at 2.12\,\micron.
    The top panel presents the centroid velocity of the emission,the middle panel shows the {full-width at zero intensity} of the line profile, and the bottom panel indicates the {terminal velocity at the blue side} of the \hh emission.
    The black contours corresponds to 10\%, 25\%, 50\% and 75\% of the peak intensity of the \hh emission.}
    \label{fig_h2_velocity}
\end{figure}

In Figure\,\ref{fig_h2_velocity} we present the velocity and the dispersion map of the \hh\,(1--0)\,S(1) transition at 2.12\,\micron, the brightest \hh transition identified in the \K-band NIFS data cube.
The radial velocity was corrected by the rest velocity of W33A ($V_{\rm LSR}$\,=\,36.7\,\kms, \citealt{Lumsden13}).
The top panel of Fig.\,\ref{fig_h2_velocity} shows that the bulk of the \hh emission is located at a radial velocity of $-$56\,\kms, showing no significant velocity gradient in the FOV.
In addition, no substantial change in the full-width at zero intensity (FWZI) of the line profile is observed towards the emitting region, exhibiting values around 110\,\kms (middle panel).
The bottom panel of Fig.\,\ref{fig_h2_velocity} shows the high-velocity of the \hh emission profile, evaluated as the terminal velocity at the blue side of the \hh profile (estimated from the FWZI parameter).
The velocity map exhibits relatively higher velocity values close to the W33A protostar ($\sim$\,$-$130\,\kms) and is roughly constant at large distances ($\sim$\,$-$113\,\kms).

\begin{figure}
	\includegraphics[width=\linewidth]{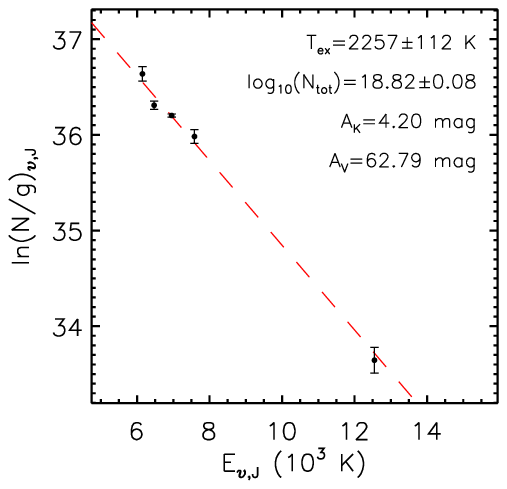} \\[-4.0ex]
    \caption{Boltzmann diagram of the \hh gas associated with W33A. The dashed line indicates the best fit to the data. The corresponding temperature, total column density and reddening values are indicated in the upper-right corner.}
    \label{fig_h2_boltzmann}
\end{figure}

We further measured the fluxes of the {individual} \hh transitions in the integrated spectrum of the {\hh cavity (shown as the green region in Fig.\,\ref{fig_W33A_nifs})} to construct a ro-vibrational Boltzmann diagram of the \hh molecule and, thus, to evaluate the physical parameters of the jet. Table\,\ref{tab:h2_fluxes} summarises the fluxes and the physical properties of each \hh transition.
Figure\,\ref{fig_h2_boltzmann} presents the Boltzmann diagram of the \hh emission, constructed using the methodology {adopted by} \citet{Caratti06,Caratti15} and briefly explained below.

\begin{table}
    \caption{H$_2$ transitions identified in the spectrum of the {cavity of the outflow}.}
    \label{tab:h2_fluxes}
    \begin{tabular}{r|rrrrrr}
    \hline
\multicolumn{1}{c}{Transition}   &   \multicolumn{1}{c}{Flux} & 
\multicolumn{1}{c}{$\sigma_F$} & 
\multicolumn{1}{c}{$\lambda$} & 
\multicolumn{1}{c}{$g$} & 
\multicolumn{1}{c}{$E_{\rm upper}$} & 
\multicolumn{1}{c}{$A$} \\
   &   \multicolumn{2}{c}{(10$^{-19}$\,erg\,cm$^{-2}$\,s$^{-1}$)} & 
\multicolumn{1}{c}{(\micron)} & 
 & 
\multicolumn{1}{c}{(K)}  & 
\multicolumn{1}{c}{(10$^{-7}$\,s$^{-1}$)}\\
    \hline
1-0 S(2) &  8.51 & 0.36 & 2.0338 &  9 &  7584 & 3.98 \\
1-0 S(1) & 30.19 & 0.40 & 2.1218 & 21 &  6956 & 3.47 \\
1-0 S(0) &  9.09 & 0.36 & 2.2235 &  5 &  6471 & 2.53 \\
2-1 S(1) &  5.16 & 0.40 & 2.2477 & 21 & 12550 & 4.98 \\
1-0 Q(1) & 58.49 & 0.44 & 2.4066 &  9 &  6149 & 4.29 \\
    \hline
    \end{tabular} \\
    \textbf{Notes:} the energy levels are from \citet{Dabrowski84}, and the Einstein coefficients are from \citet{Turner77}.
\end{table}   

The upper-level column density of each \hh transition ($N_{\upsilon,J}$, where $\upsilon$ and $J$ denotes the upper vibrational and rotational levels, respectively) is evaluated as:
\begin{equation}
    N_{\upsilon,J} = \frac{4 \pi \Omega}{h} \frac{F_{\upsilon,J}}{A_{\upsilon,J} \nu_{\upsilon,J} g_{\upsilon,J}}
\end{equation}
\noindent where $h$ is the Planck constant (in erg\,s), $\Omega$ is the solid angle of the emitting region (in sr), $F_{\upsilon,J}$ is the integrated intensity of the transition (in erg\,cm$^{-2}$\,s$^{-1}$\,sr$^{-1}$),  $A_{\upsilon,J}$ corresponds to the Einstein coefficient of the transition (in s$^{-1}$),  $\nu$ is the frequency (in Hz), and $g_{\upsilon, J}$ is the statistical weight of the transition.
The upper-level column density of each \hh transition is plotted against their upper-level energy ($E_{\upsilon,J}$, in K).
Assuming the gas is in Local Thermodynamic Equilibrium (i.e. follows the Boltzman distribution), a linear fit to the data in a log-normal scale (i.e. $\log(y)$\,=\,$\alpha\,x + \beta$) provides an estimate of the excitation temperature (\tex=\,$-1/\alpha$), and the total column density [\ntot=\,$Q(T_{\rm{ex}}) \exp(\beta)$] of the \hh gas.
The \K-band extinction (\Ak) can also be inferred by choosing the best \Ak value that maximises the correlation between the quantities shown in the Boltzmann diagram (for details, see \citealt{Caratti15}).

    The best-fit to the Boltzmann diagram shows that the molecular gas is associated with a \tex value of (2.3$\pm$0.1)$\cdot$10$^3$\,K, and a total \hh column density of $\log$(\ntot/cm$^{-2}$)\,=\,18.8$\pm$0.1\,dex.
    The \K-band reddening is about 4.20$\pm$0.1\,mag, which is equivalent to an $A_{V}$ of 63$\pm$2\,mag assuming the $A_{\rm V}/A_{\rm K}$ ratio of 14.95 from \citet{Damineli16}.
 
    The origin of the \hh emission can be inferred using the ratio of \hh transitions such as the (1--0)\,S(1)/(2--1)\,S(1). Based on the fluxes reported in Table\,\ref{tab:h2_fluxes} and correcting them for reddening, we obtained a (1--0)\,S(1)/(2--1)\,S(1) ratio of 9.3\,$\pm$\,1.4. The observed ratio is close to the theoretical value of $\sim$10 for thermal excitation of the \hh emission through shocks, and significantly larger than the predicted value of 1.8 for the case of radiative excitation through UV photons from the protostar \citep{Black76}.
    
\begin{figure*}
	\includegraphics[width=0.75\linewidth]{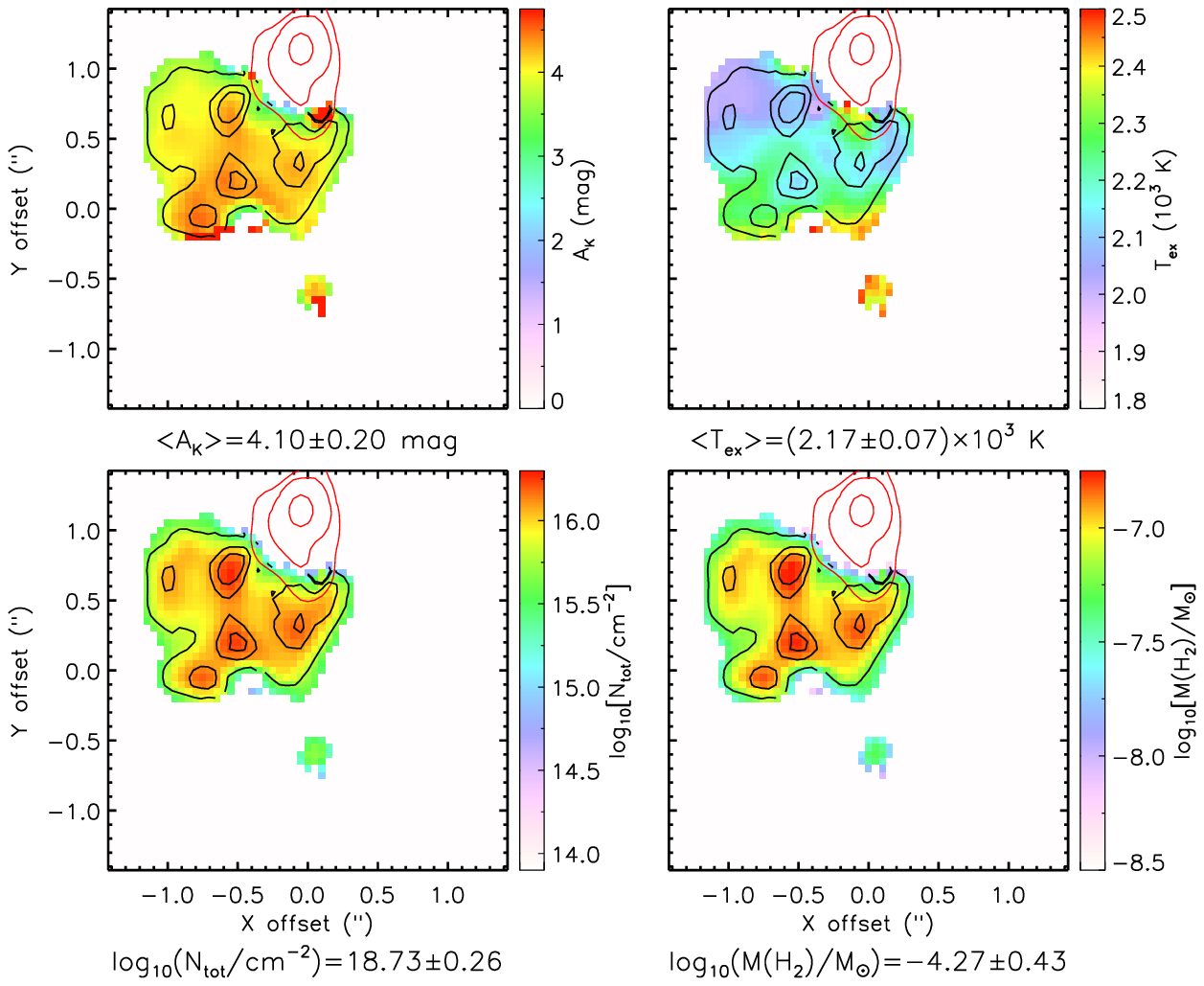} \\[-1.0ex]
    \caption{Pixel-to-pixel derivation of the \hh parameters {for those pixels exhibiting fluxes above a 3-$\sigma$ threshold}. From top left to bottom right: the \Ak, \tex, column density and mass of the \hh gas. The black contours indicates the integrated H$_2$\,(1--0)\,S(1) flux. The red contours indicate the position of the point-like source and its extended continuum emission. The range of each parameter was chosen in order to improve the visualisation of the maps. The mean value of the \Ak and \tex, and the total column density and mass of the \hh gas are shown at the bottom of their respective panels.}
    \label{fig_h2_p2p}
\end{figure*}

    By taking advantage of the spatial information available in the NIFS data, we implemented a pixel-to-pixel evaluation of the physical parameters of the \hh gas, {as presented in Fig.\,\ref{fig_h2_p2p}.}
    Spatial maps of \Ak, \tex and \ntot were obtained by constructing and fitting the Boltzmann diagram of the \hh transitions for {the spaxels exhibiting fluxes above a 3-$\sigma$ threshold for all the five \hh transitions listed in Table\,\ref{tab:h2_fluxes}}.
    In addition, the mass of the \hh gas ($M_{H2}$) was computed  based on the total column density of the \hh molecules as:
\begin{equation}
    M_{\rm H2}(x,y) = 2 \mu m_H A_{\rm pix} N_{\rm tot}(x,y)
\label{eq_mass_h2}
\end{equation}
\noindent where the (x,y) indices correspond to the position of a given spaxel of the data cube, $\mu$\,=\,1.24 is the mean atomic weight of the \hh, $m_{\rm H}$ is the proton mass (in g) and $A_{\rm pix}$ corresponds to the projected area of the spaxel (in cm$^2$).

    The top left panel of Fig.\,\ref{fig_h2_p2p} presents the distribution of the \K-band extinction over the {\hh emission associated with the cavity of the outflow}.
    The median \K-band extinction of the \hh emission is \Ak=\,4.1$\pm$0.2\,mag, corresponding to an $A_{\rm{V}}$ of $\sim$61\,$\pm$\,3\,mag \citep{Damineli16}.
    
    The map of the excitation temperature of the gas is presented in the top right panel of Fig.\,\ref{fig_h2_p2p}. The median \tex value is (2.2\,$\pm$\,0.1)\,$\cdot$\,10$^3$\,K.
    The total column density of the \hh gas is presented in the bottom left panel.
    The distribution of \ntot is similar to the flux of the H$_2$\,(1--0)\,S(1) line at 2.12\,{\micron}, indicated by the contours, suggesting a linear relationship between the brightness of this transition and the logarithm of the column density. The total column density of the gas is about {$\log$(\ntot)$\sim$\,18.7$\pm$0.3\,dex}.
    The bottom right panel of Fig\,\ref{fig_h2_p2p} shows the mass of the \hh gas, computed using Eq.\,\eqref{eq_mass_h2}. The total mass of the emitting \hh {region} is about {(5.4$\pm$0.1)\,$\cdot$\,10$^{-5}$\,\msun}.
    
    A rough estimate of the dynamical timescale of the {cavity of the outflow, probed by the \hh emission,} can be evaluated as
    \begin{equation}
        t_{\rm dyn} = \frac{\ell_{\rm H2}}{|v_{\rm H2}|} = \frac{\cos{(i)}}{\sin(i)} \frac{\ell_{\rm proj}}{|v_{\rm rad}|} \approx \frac{4.74}{ \tan(i)} \frac{\ell_{\rm proj}\,(\rm{AU})}{|v_{\rm rad}|\,(\rm{km\,s^{-1}})}\,\rm{yr}
        \label{eq_tdyn}
    \end{equation}
    \noindent where $\ell_{\rm proj}$ is the projected length of the {emitting region} in the plane of the sky ($\ell_{\rm proj}$ = $\ell_{\rm H2} \cdot \sin(i)$), $v_{\rm rad}$ is its {radial} velocity{ of the emission ($v_{\rm rad} = v_{\rm H2} \cdot \cos(i)$), and $i$ is the inclination angle of the structure ($i$\,=\,30$^\circ$)}.
    %
    The {\hh emission} extends up to 1\farcs65 from the driving source, corresponding to a {projected} linear scale of 3960\,AU at the distance of 2.4\,kpc.
    The emission peaks at 2.1216\,{\micron}, corresponding to a velocity of about $-$28.3\,\kms.
    Considering the rest velocity of W33A, $v_{lsr}$\,=\,36.7\,\kms \citep{Lumsden13}, the projected {radial} velocity of the {\hh emission} is about $-$65\,\kms, {in agreement with the analysis presented in Fig.\,\ref{fig_h2_velocity}}.
    Using Eq.\,\eqref{eq_tdyn} {and propagating the error on each input parameters\footnote{We considered a typical $\pm$20$^\circ$ on the inclination angle, $\pm$29\,\kms on the velocity (half of the velocity resolution of the NIFS data), and an error of 10\% on the projected length}, we obtained} a dynamic timescale of about {(5.0\,$\pm$\,2.4)$\cdot$10$^2$\,years}, suggesting the \hh {emission arises from} a relatively young structure. {We further obtained a rough estimate of the mass-loss rate of the ejected gas as $\dot{M}_{\rm H2}$\,=\,$M_{\rm H2}$/$t_{\rm dyn}$\,$\sim$\,{(1.1$\pm$0.5)\,$\cdot$\,10$^{-7}$\,\msun\,yr$^{-1}$}.}
    
\subsubsection{Spectro-astrometry of the Brackett-Gamma feature}
\label{sec_spectroastrometry}

According to the PCA results, the \brg emission is observed towards the W33A protostar (PC1) and the extended nebulosity to the south (PC1 and PC2), the rotating disc associated with W33A (PC3), and compact emitting regions closer to the W33A protostar (PC4 and PC5).

\citet{Davies10} reported that the \brg emission was probing the {cavity of the large-scale \K-band outflow} at very small scales using the spectro-astrometry of the \brg line, a technique that combines the spectral and spatial information of the data cube to resolve kinematic structures at sub-pixel scales with impressive accuracy (i.e. one tenth of a milliarcsecond).
Thus, we adopted the same methodology presented by \Davies to check if the post-processing of the NIFS data led to a significant improvement of the data at sub-pixel scales.

\begin{figure*}
	\includegraphics[width=\linewidth]{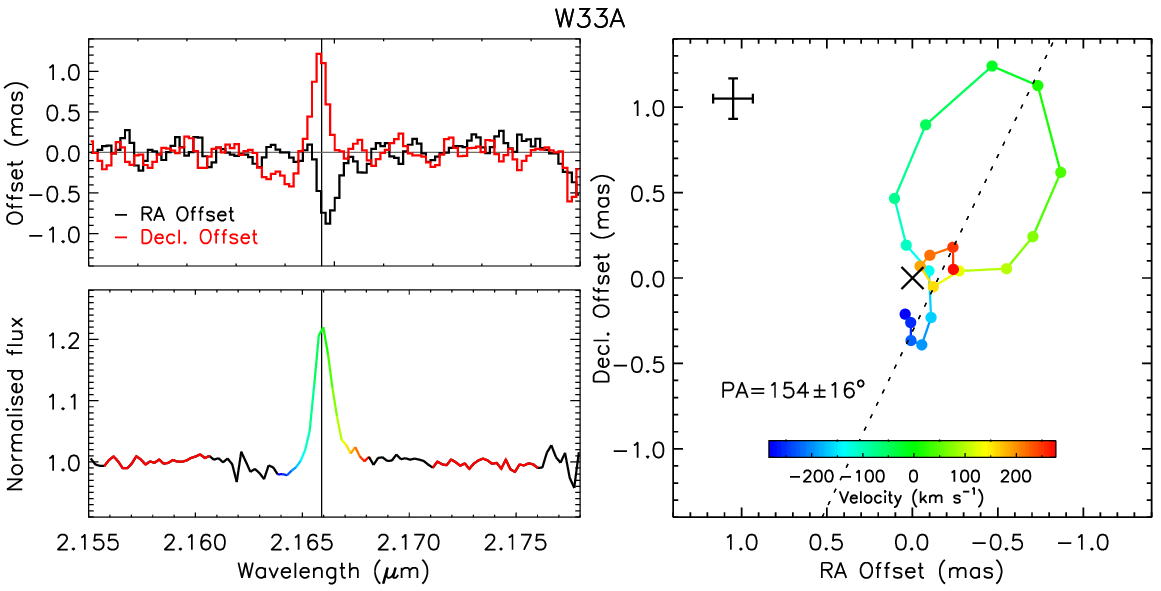}  \\[-1.0ex]
    \caption{Spectro-astrometry of the \brg feature.
    Top left panel: the centroid position in the right ascension (RA, in black) and declination axis (Decl., in red) as a function of the wavelength. The baseline is indicated by a horizontal grey line. The peak of the \brg feature (at 0\,\kms) is indicated by a vertical line.
    Bottom left: the normalised \brg flux. The coloured region indicates the spectral interval shown on the right panel. {The spectral regions used to estimate the positional error of the centroid (about $\sim$0.12\,mas in each direction) are indicated in red}. Right panel: Spatial variation of the \brg centroid position. The velocity with respect to the peak of the emission is indicated by the horizontal colour bar shown in the bottom region of the plot. The positional error 
    is shown in the left top corner of the plot. The dashed line indicates the position-angle (PA) of the red and blue high-velocity components.}
    \label{fig_brg_spectroastrometry}
\end{figure*}

The spectro-astrometry of the \brg feature of W33A is presented in Fig.\,\ref{fig_brg_spectroastrometry}.
The left panel presents the offset of the centroid in the right ascension (black curve) and declination axes (red curve) as a function of the wavelength (top panel), together with the \brg flux at the same wavelength scale (bottom). 
The right panel presents the spatial variation of the centroid position as a function of the velocity, matching the colour scale of the \brg profile shown in the left panel.
{The positional error of 0.12\,mas, indicated in the top left corner, was based on the centroid position offset of the nearby continuum around the \brg line (corresponding to the spectral regions shown in red).}

The blue and red high-velocity components (corresponding to velocities of $-$300 to +300\,\kms) are offset by roughly $\sim$\,0.55\,mas, from the south-east to the north-west direction, corresponding to $\approx$\,1.3\,AU at the distance of 2.4\,kpc.
The displacement between the blue and red high-velocity components roughly follows a linear distribution with a position angle (PA) of {154$\pm$26$^\circ$ (from N to E)}, which is compatible with the results reported by \Davies {(156$\pm$25$^\circ$\footnote{{equivalent to the value reported  by the authors, PA\,=\,113$\pm$25$^\circ$, oriented from E to N}})}.

\subsubsection{Kinematics of the CO absorption features}
\label{sec_co_ppfx}

    We used the {\it Penalized Pixel-Fitting} algorithm \citep[pPXF][]{Cappellari04} to derive the velocity map of the CO absorption features observed in the spectrum of W33A. 
    The pPXF uses a template spectrum to fit and evaluate the radial velocity at each spaxel in the data cube. The procedure starts with an initial guess of the velocity and width parameters and solves a Gauss-Hermite function based on the template spectrum to fit the observed one. The residuals between the model and the observed spectra are perturbed using a penalised function and the best model is chosen based on a $\chi^2$ minimisation procedure, delivering a set of parameters (V, $\sigma$, $h_3$, ... $h_m$) and their uncertainties for each spaxel of the datacube. For this work, we focused the further analysis on the interpretation of the radial velocity measurements only.
    The brightest spaxel of the continuum emission at $\sim$2.3\,{\micron} (at the central position of the point-source object) was adopted as the template spectrum (corresponding to the rest-velocity at 0\,\kms) to derive the kinematics of the CO absorption lines {using the spectral region ranging from 2.32 to 2.37\,{\micron}, containing both the low-J CO features of the R and the P branches}.
    To spatially sample the rotating structure, the calculations were performed only for the spaxels within a 0\farcs5 radii around the central position of W33A.
    
\begin{figure}
	\includegraphics[width=0.9\linewidth]{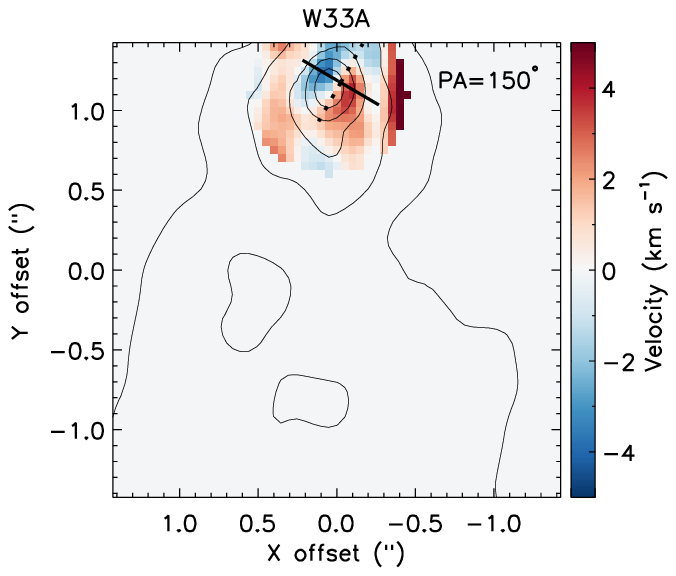} \\[-4.0ex]
	\includegraphics[width=0.9\linewidth]{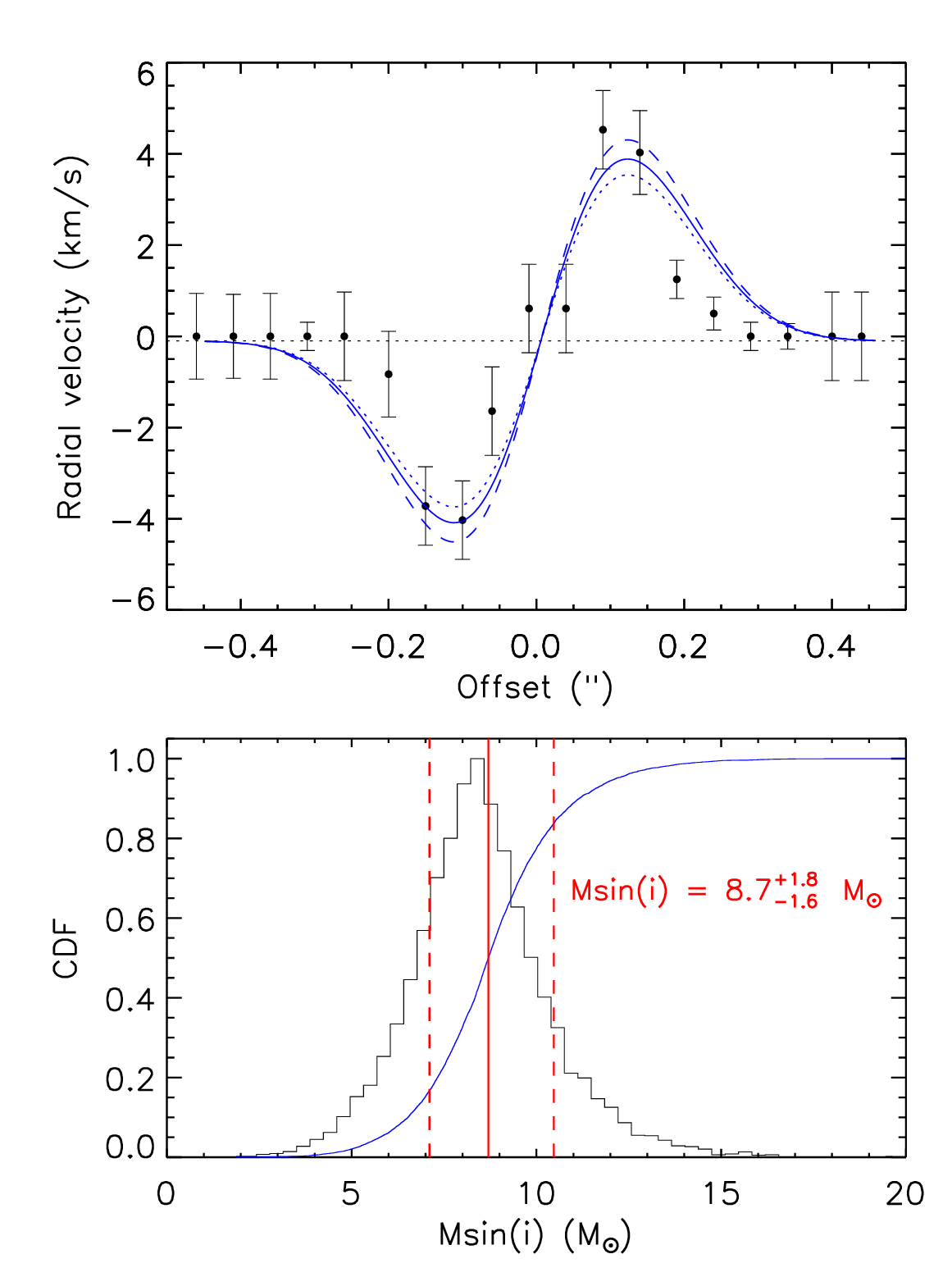}  \\[-6.0ex]
    \caption{Top panel: Radial velocity map of the CO absorption features towards W33A, derived with the pPXF fitting.
   The black contours are placed at the 3$^{n}-\sigma$ level ($n$\,=\,$0,1,2,...$) of the integrated continuum emission at $\sim$2.3\,{\micron}.
   The velocity scale is presented on the right side of the map.
   The dashed black line indicates the position angle of 150$^\circ$ (from N to E) of the rotation axis of the disc, following the velocity gradient between the blue- and the red-shifted disc components.
   The filled black line is perpendicular to the disc axis, indicating the direction where the position-velocity diagram was extracted.
   Middle: The position-velocity diagram for the CO absorption features. The black circles and their errors corresponds the measurements using the pPXF analysis. The best Keplerian rotation curve for $M \sin(i)=8.7$\,\msun is indicated by the solid blue line. The dotted and dashed blue lines indicate the Keplerian curves considering the upper and lower 1-$\sigma$ errors on the mass, respectively.
   Bottom panel: The cumulative distribution function (CDF) of the $M\sin(i)$ value evaluated for 10,000 Monte-Carlo simulations is shown as the solid blue curve.
   The relative histogram of $M \sin(i)$ values is shown as the solid black line. The solid vertical line corresponds to the median value and the dashed lines are placed at the 16 and 84\% confidence values, corresponding to the 1-$\sigma$ error bars.}
    \label{fig_co_kin_W33A}
\end{figure}

    Figure\,\ref{fig_co_kin_W33A} presents the results of the pPXF analysis. The left panel indicates the map of radial velocities overlaid with contours of the continuum emission.
    The map indicates a clear gradient between negative and positive radial velocities {very similar to the pattern observed in the tomogram of PC3 (see Figs.\,\ref{fig_pca_final_PC3} and \ref{fig_PC3_cut})}, suggesting that the kinematics arise from a rotating disc as reported by \Davies.
    The rotating axis of the structure is aligned with a position angle of about {150\,$\pm$\,10$^\circ$} (from N to E), {compatible with the orientation of the rotating axis observed in the tomogram of PC3 (see Fig.\,\ref{fig_PC3_cut}).}
    The velocity gradient arises from a relatively compact region of $\sim$0\farcs5, corresponding to a projected length of about 1,200\,AU at the distance of 2.4\,kpc.
    
    {We used the velocity map delivered by the pPXF analysis to extract a position-velocity (PV) diagram perpendicularly to the rotation axis of the disc. The results are presented in the middle panel of Fig.\,\ref{fig_co_kin_W33A}, exhibiting the mean value and the standard deviation of the velocity at each position, evaluated within 3-pixel wide bins across the extraction axis (indicated by the filled black line in the top panel of Fig.\,\ref{fig_co_kin_W33A}).}
    The PV diagram indicates a clear separation between the blue- and red-shifted components (with peak velocities around $+$\,5 and $-$\,4\,\kms, respectively) separated by roughly 0\farcs2.
    The best Keplerian model, shown as the solid blue curve, indicates that the total mass of the ``protostar+disc'' system is about 8.7$^{+1.8}_{-1.6}$\,\msun. We note that the velocity falls more rapidly than any Keplerian model at larger radii. 
    The mass and its standard deviation were evaluated using a Monte-Carlo method (see results in the bottom panel of Fig.\,\ref{fig_co_kin_W33A}), varying the input data randomly within their errors. We used a total of 10,000 simulations to achieve the reported values.
    Assuming the inclination angle of $i$\,=\,60$^\circ$ adopted by \Davies \citep[estimated from interferometric mid-IR observations from][]{deWit10}, the inclination-corrected mass of the ``protostar+disc'' system in W33A is 10.0$^{+2.1}_{-1.8}$\,\msun. 

\section{Discussion}
\label{sec_discussion}

This work presents a re-analysis of the $K$-band IFU observations of the well-known high-mass protostar W33A with an eye to further extend the analysis presented originally by Davies et al. (\Davies) through application of a Principal Components Analysis.

First, we performed the flux-calibration of the data cube followed by the differential atmospheric refraction (DAR) correction. {The DAR effects correspond to a slight positional offset of the source as a function of the wavelength (see  Fig.\,\ref{fig_atmrefraction}).
The proper correction of such effects is necessary to properly compare the spatial information at distinct wavelengths in the data cube, and to perform measurements at a very small fraction of a pixel as presented in this work; see also \Davies.}

Then, we performed the PCA tomography of the data cube, and found that most of the astrophysical information of the data cube is well-explained by the first five Principal Components (PCs):
PC1 shows that the embedded compact source associated with W33A dominates the emission in the \K-band;
PC2 presents the differences between the compact and the extended emission regions, mostly showing an anti-correlation introduced by reddening effects;
PC3 exhibits the kinematics of a rotating circumstellar disc around W33A, probed by the low-$J$ CO absorption lines and exhibiting a contribution of ionised gas;
PC4 probes the cavity of the jets, showing a roughly perpendicular structure aligned to the rotating axis of the disc, which is likely reflecting the emission of the inner region of the disc and the protostar itself; and
PC5 exhibits the anti-correlation between the {extended \hh emission} and the \brg and CO emission arising from the compact object.

\begin{figure}
\centering
    \includegraphics[width=\columnwidth]{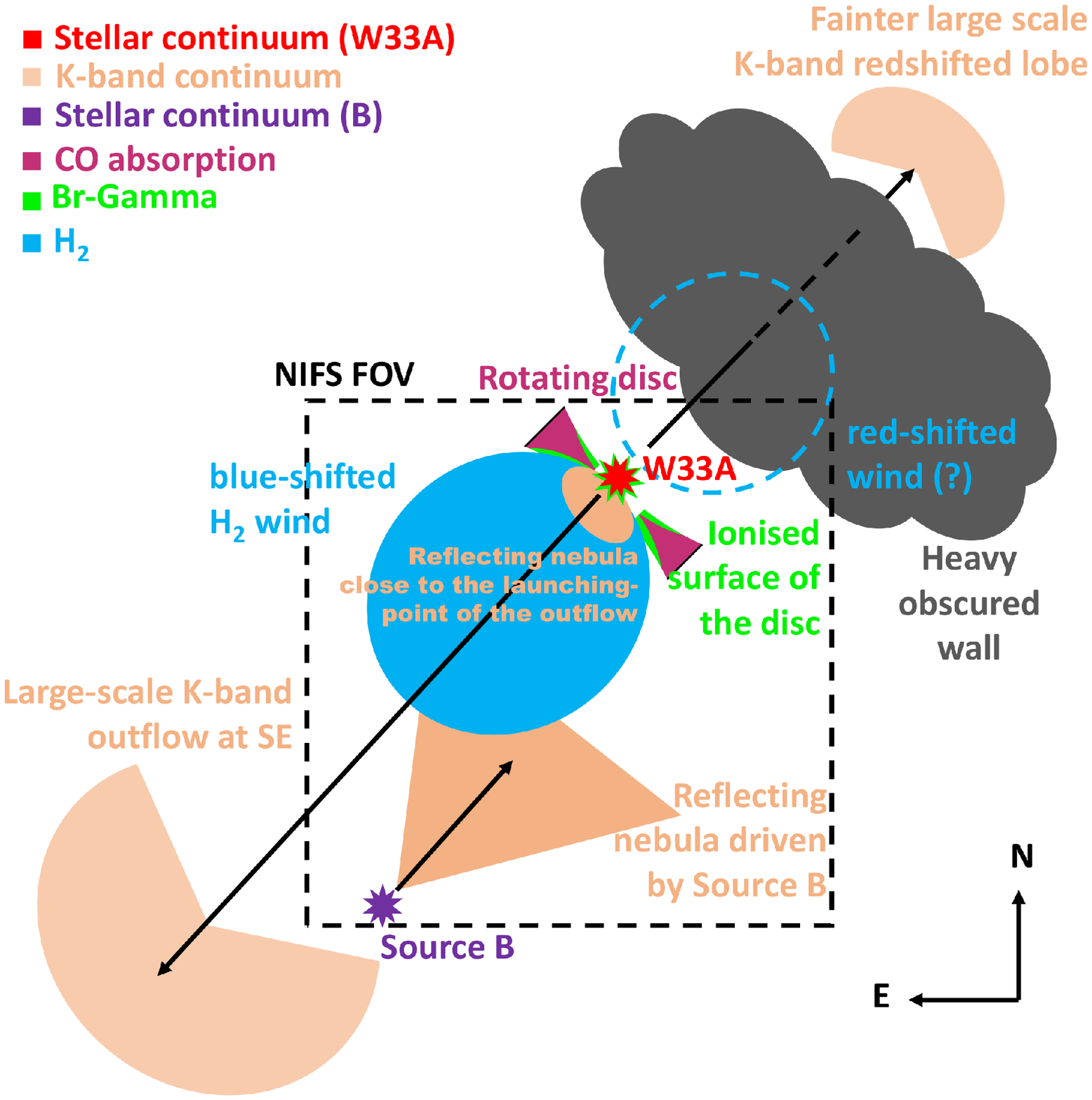}
    \caption{Interpretation of the near-infrared $JHK$ maps and NIFS \K-band observations of W33A. The illustration roughly follows the same orientation of the observations (North is up and East is to the left). The structures are labelled and shown in different colours, as indicated in the top left corner.}
    \label{fig_model_interpretation}
\end{figure}

Along with the PCA results, the analysis of the near-infrared $JHK$ maps of W33A allowed us to obtain an overall interpretation of the environment associated with the protostar and its surroundings, as illustrated in Fig.\,\ref{fig_model_interpretation}.

The data cube was reconstructed using the first five PCs, suppressing a considerable fraction of noise  associated with PC6 and higher-orders PCs (see Scree test in Fig.\,\ref{fig_screetest}).
Then, we followed a similar analysis as that performed by \Davies to validate our results using the reconstructed data cube. In general, we found that the post-processing routines and noise suppression of the data cube led a significant improvement of the results, as discussed below. 

\subsection{Multiple IR sources within W33A}

The comparison of the large-scale NIR maps with the NIFS FOV in Fig.s\,\ref{fig_nir_w33a} and \ref{fig_nifs_jhk} shows that different sources are likely contributing to the observed extended \K-band emission detected in the field.

In Fig.\,\ref{fig_pca_final_PC2}, PC2 {shows} the anti-correlation between the circumstellar region of W33A (probed mostly by the resolved CO absorption features) and the extended emission mostly probed by the continuum and \brg emission.
The tomogram of PC2 exhibits a clear separation between the extended emission closer to W33A and the larger structure towards the S.

A point-like source is observed at the SE edge of the NIFS FOV, consistent with the position of the Source\,B (see Fig.\,\ref{fig_nir_w33a}).
At the distance of W33A, the linear separation of W33A and Source\,B is about $\sim$7000\,AU. In addition, the detection of a large-scale bipolar jet in the E-W direction associated with the $H$-band compact Source\,C (see Fig.\,\ref{fig_nir_w33a}) indicates that multiple sources are located within this region.
The $JHK$ photometry and the colour-magnitude and colour-colour diagrams presented in Fig.\,\ref{fig_cmd_w33a} allowed us to confirm that Source\,B and C are compatible with young, protostellar objects.

\subsection{The mass of W33A and the compact circumstellar disc}
\label{sec_massCO}

We obtained a robust estimate of the total mass of the ``disc+protostar'' system and its error based on the pPXF analysis of the CO absorption features at 2.35\,{\micron} (see Fig.\,\ref{fig_co_kin_W33A}).
The best fit of the position-velocity diagram is consistent with a Keplerian rotation structure around a total mass of M$\sin(i)$\,=\,8.7$^{+1.8}_{-1.6}$\,\msun, assuming a distance of 2.4\,kpc.
Assuming the inclination angle of 60$^\circ$ as adopted by \Davies, the mass is 10.0$^{+2.1}_{-1.8}$\,\msun.
If considering a typical $\pm$20$^\circ$ uncertainty on the inclination angle, the upper-limit of the mass increases by a factor of two, 10.0$^{+4.1}_{-2.2}$\,\msun, {indicating that the inclination uncertainty is a significant source of error on the total mass.}
The inclination angle value of 60$^\circ$ adopted in \Davies and in this work was derived from the analysis of interferometric mid-IR observations from \citet{deWit10}. 
Those authors argued that the derived inclination angle of the system is consistent with the fact that W33A exhibits no strong $H-$band emission arising from the central star (as one would expect for a less inclined system), and it is also in agreement with the relatively weak large-scale \K-band redshifted outflow lobe (Fig.\,\ref{fig_nir_w33a}).
\citet{Ilee13} reported a more accurate determination of the inclination angle of the W33A system as $i(^\circ)$\,=\,37$^{+16}_{-11}$ based on the modelling of the CO bandhead emission at 2.3\,{\micron} using high spectral resolving power observations (R$\sim$30,000).
More recently, \citet{Fedriani20} analysed the \K-band emission of the high-mass YSO IRAS\,11101$-$5829. Those authors reported that the CO bandhead emission exhibits low velocities, likely arising from a face-on rotating disc (i.e. high inclination angle). However, the geometry of \hh the jet driven by the IRAS source requires that the disc must be roughly edge-on, with low inclination angle. They conclude that most of the CO emission is observed as reflected and scattered light in the outflow cavity walls, and therefore the inclination of the system can be incorrectly interpreted based on the modelling of the CO bandhead emission.
In addition, the identification of the low-$J$ CO features in absorption in the spectrum of W33A (e.g. see Fig.\,\ref{fig_W33A_nifs_original}) is also indicative of an inclined disc and, therefore, not compatible with such a small inclination value as that reported by \citet{Ilee13}. 

\citet{Davies10} derived the total mass of W33A as 15$^{+5}_{-3}$\msun, considering a distance of 3.8\,kpc.
Scaling their mass to the revised distance adopted in our work, 2.4\,kpc, the total mass of W33A is 9.5$^{+3.2}_{-1.9}$\,\msun, 
which is in good agreement with the mass we obtained with the pPXF analysis.
The pPXF method allows the determination of the kinematics of very faint spectral features, such as the CO absorption features, and also provides a robust estimate of the errors, an important extension of the \Davies result.

The rotation curve based on the CO absorption lines reported by \Davies was derived by smoothing the original CO absorption velocity centroids using a 5-pixel boxcar filtering. This procedure suggests this component is a relatively extended structure with an angular size of about {1\arcsec} around the source (see their Fig.\,14), corresponding to a radius of $\sim$2,000\,AU from the protostar.
Our results show a slightly more ordered view of the absorbing gas and indicate that the velocities rapidly decrease for radii larger than 0\farcs1 ($\sim$500\,AU) from the central position.
    The velocities deviate from Keplerian outside the inner 0\farcs2 region of the PV diagram shown in Fig.\,\ref{fig_co_kin_W33A} (corresponding to a projected length of $\sim$500\,AU at the distance of the source).
    These results suggest the circumstellar disc associated with W33A is a compact structure, similar to the one observed towards the high-mass YSO IRAS\,13481-6124 by \citet{Kraus10}. Based on interferometric infrared observations, those authors reported a $\sim$130\,AU flared disc around a 18\,\msun protostellar object lying within the IRAS source.
    
    Recently, \citet{Maud17} presented the analysis of interferometric ALMA observations of the W33A protostar at Band\,6 and 7 (230 and 345 GHz, respectively) with angular resolutions of $\sim$0\farcs2, similar to that of the NIFS data (0\farcs15). Those authors have found that the kinematic signatures associated with W33A are inconsistent with a Keplerian disc at the angular resolution of their data. However, they did not exclude the possibility of an underlying compact disc with Keplerian rotation at smaller scales, such as confirmed by the NIFS \K-band IFU observations.
    ALMA observations obtained at larger baseline lengths (e.g. 16\,km) are, therefore, necessary to disentangle the structure and kinematics of the innermost region of W33A.

\subsection{A compact wind traced by the molecular hydrogen gas}

The near-IR maps from Fig.\,\ref{fig_nir_w33a} shows no evidence for a large-scale \hh jet associated with W33A, as one would expect for an active protostar (e.g. \citealt{Varricatt10,Navarete15}).
{At longer wavelengths, recent} ALMA observations of W33A \citep{Maud17} have shown that high-velocity emission of CO and SiO lines are probing a relatively compact jet-like structure with angular sizes of $\sim$0\farcs5.
Despite of that, no evidence for a collimated jet-like structure was observed in the NIFS FOV, but a wide-angle \hh polar emission driven by the protostar. 
The nature of the \hh emission can be traced by the other structures associated with W33A. The orientation of the sub-pixel structure probed by the spectro-astrometry of the \brg feature (PA\,=\,154$\pm$26$^\circ$, see Fig.\,\ref{fig_brg_spectroastrometry}) is compatible with the disc rotation axis (150$\pm$10$^\circ$, see Fig.\,\ref{fig_co_kin_W33A}), suggesting that the ionised Hydrogen is tracing the base of the jet/wind in high-mass YSOs as also seen in IRAS\,13481-6124 through near-IR interferometry \citep{Caratti16}.
In addition, the orientation of the \hh emission (PA\,$\sim$\,145$^\circ$, see Fig.\,\ref{fig_pca_final_PC5}) is also compatible with the ionised gas and the disc axis, favouring the interpretation of a wide wind traced by the \hh emission rather than the emission of scattered light in the cavity of the large-scale outflow.
This interpretation is also in agreement with the physical parameters of the \hh emission derived through the \hh (1--0)\,S(1)/(2--1)\,S(1) line ratio and the ro-vibrational analysis presented in Sect.\,\ref{sec_h2jet}, confirming the thermal origin of the emission.

The physical parameters of {the \hh emission} 
were derived by constructing the Boltzmann diagram of the {different \hh transitions of the entire \hh emitting region  (see Fig.\,\ref{fig_h2_boltzmann})}, and also by evaluating the parameters for each spaxel in the data cube (see Fig.\,\ref{fig_h2_p2p}).
Based on the pixel-to-pixel analysis, the {\hh emission exhibits} a mean \Ak reddening of 4.3\,$\pm$\,0.2\,mag, an excitation temperature of about 2.2$\times$10$^3$\,K, and a total \hh column density of 10$^{18.7\pm0.2}$\,cm$^{-2}$. The total mass of the {\hh emission} was estimated to be M$_{\rm H_2}$\,=\,(4.6$\pm$1.0)\,$\cdot$\,10$^{-5}$\,\msun.
In general, there is a good agreement between the physical parameters evaluated from both methods (i.e. the integrated spectrum and the pixel-to-pixel analysis).
The spatially resolved analysis also shows the clumpy character of the \hh emission.
The \Ak map shown in the top left panel of Fig.\,\ref{fig_h2_p2p} indicates the extinction towards the \hh jet is strong and variable, with values ranging from $\sim$1.5 up to 5.0\,magnitudes in the \K-band.
The \Ak values are larger towards the positions where the \hh emission is stronger indicating dense clumps of gas and dust perhaps along the cavity walls of the large-scale outflow (recall the emission is consistent with shocked gas). But the conclusion is that the \hh is probing a wind.

We further compared the physical parameters of the {\hh emission} of W33A using the typical values observed toward large-scale \hh jets of high-mass protostars from \citet{Caratti15}.
The total column density of the {\hh wind driven by W33A} is consistent with the range of column density values (10$^{17}$ to 10$^{20}$\,cm$^{-2}$) reported by those authors.
    In addition, they also derived the mean visual extinction ($A_{V}$) of the \hh jets ranging from 1 to 50\,mag, with a median of $A_{V}$\,=\,15\,mag.
    The upper limit of the {visual extinction values reported by them is compatible} with the median $A_{V}$ value for the {\hh wind} associated with W33A, $A_{V}$\,=\,63$\pm$2\,mag.

Finally, we cannot infer the \Ak towards the W33A protostar based on the \hh analysis due to the absence of molecular gas emission closer to the protostar.
    Alternatively, the \Ak value of the point-like source can be evaluated based on its \K-band excess ($E_{H-K}$). Assuming the extinction law from \citet{Damineli16}, the \Ak is obtained as $A_{K} = 1.3 \cdot E_{H-K}$. For W33A, the observed $H-Ks$ colour index is about 4.0\,mag.
    Considering the intrinsic $H-Ks$ colour index for a B2\,V star ($\sim$10\,\msun), $(H-Ks)_0$\,=\,$-$0.05\,mag \citep{Wegner03}, we obtain an extinction in the $K_s$-band of $A_{Ks}$\,$\approx$\,5.3\,mag for the point-like source (or $A_{V}$\,$\approx$\,80\,mag). This value is in agreement with the upper-limit of the \Ak values associated with the \hh {emission}, shown in Fig.\,\ref{fig_h2_p2p}.
    
\subsection{Different structures associated with ionised gas in W33A}
\label{sec_discussion_brg}

The \brg emission arises from gas at typical electron temperatures of 8,000-10,000\,K and it is observed in distinct structures of W33A, suggesting that the hydrogen recombination plays a significant role in cooling the gas. The Principal Component 1 (Fig.\,\ref{fig_pca_final_PC1}) indicates the bulk of the \brg emission arises at the central source of W33A.
The source of ionised gas could be an unresolved ultra-compact $\rm{H\,\textsc{ii}}$ region around the protostar, in agreement with the absence of 5\,GHz radio continuum emission towards the protostar \citep{Purcell13}.

In Principal Component 2 (Fig.\,\ref{fig_pca_final_PC2}), the \brg emission mostly arises from the the extended emission observed in the FOV, either originated by Source\,B or W33A itself.

In PC3 (Fig.\,\ref{fig_pca_final_PC3}), the \brg feature exhibits a similar kinematic pattern as the CO absorption features.
Our interpretation is that the CO features are probing a compact and rotating circumstellar disc around W33A (see Fig.\,\ref{fig_co_kin_W33A}), and the surface of the disc is partially ionised, producing a disc wind, similar to that observed in Herbig Ae/Be stars \citep{Tambovtseva16} but perhaps reaching larger radii owing to the more massive central object.
Although the analysis of the \hh emission does not indicate the presence of any photo-ionised region in the extended wind, we cannot rule out the presence of photo-evaporating wind in $\sim$100\,AU scales closer to the surface of the underlying cold disc, where the \hh emission is weak probably because the molecules are being dissociated. As the wind moves away from the central region, the \hh molecules are excited but not dissociated, producing the observed extended structure as observed in e.g. Fig.\,\ref{fig_h2_p2p}.
This interpretation is complementary to the spectro-astrometric result shown in Sect.\,\ref{sec_spectroastrometry}, which shows the ionised gas tracing the cavity of the outflow at milli-arcsecond scales, while PC3 shows evidences for the presence of ionised gas at $\sim$100\,AU scales, in the outer regions of the circumstellar disc.
The NIFS resolution is not sufficient to disentangle the likely ionised surface from the underlying cold disc probed by the CO features. Thus, as we pointed out above in Sect.\,\ref{sec_massCO}, follow-up observations at higher spatial resolutions (e.g. using the very long baseline of the ALMA array) focusing on the investigation of the circumstellar environment may help provide a better and more accurate view of the structure on small scales.

The PC4 in Fig.\,\ref{fig_pca_final_PC4} shows that strong \brg emission is observed very close to the W33A protostar, while the spectro-astrometry of the \brg feature (Fig.\,\ref{fig_brg_spectroastrometry}) shows the innermost region of the \hh jet cavity at milli-arcsecond scales ($\sim$1.3\,AU), with a Position Angle of 115$^\circ\pm$26$^\circ$.
Our results are in agreement with those reported by \Davies, suggesting that the spectro-astrometry of the \brg feature is likely probing the base of the bipolar wind/jet traced by the \hh emission.
{We also observe that the structure identified in Fig.\,\ref{fig_brg_spectroastrometry} is less affected by noise than that presented in \Davies, indicating that the post-processing of the NIFS data cubes and noise suppression plays a significant role on the analysis of sub-pixel scale structures within the data.}

\section{Conclusions}
\label{sec_summary}

We presented a reanalysis of the $K$-band NIFS/Gemini North observations of the protostar W33A, first done by \Davies, based on Principal Component Analysis Tomography. The PCS technique is well adapted to the analysis of complex data sets like the one analysed here. 

\begin{enumerate}
\item The PCA tomography was able to recover the spectral and spatial information of structures associated with the three key-ingredients of the disc-mediated accretion scenario of the high-mass star formation process:
the cavity of the {large-scale outflow} probed by the ionised hydrogen,
a rotating disc-like structure probed by the CO absorption features, {and a wide-angle wind traced by the \hh emission}.
All these structures were present in the first five Principal Components of the PCA Tomography.
In addition, the technique also reveals structures with different kinematics (e.g. the rotation of the disc and the extended wind structure) in a much more efficient way than investigating each spectral feature using traditional methods.

\item The ionised gas (\brg) contributes to every structure associated with the W33A protostar: although the bulk of the ionised gas is observed at the central source as expected for a high-luminosity protostar, it also tracks the inner region of the jet and its cavity, as well as the surface of the disc.
\end{enumerate}

{The comparison of the structures identified in the NIFS FOV with large-scale near-infrared maps around W33A led us to the following conclusions:}

\begin{enumerate}
  \setcounter{enumi}{2}
\item {The propagation of the \hh jet identified in the NIFS FOV is consistent with the orientation of the $K$-band reflection nebula identified in the $JHK$ maps. However, no evidence for a large-scale \hh jet driven by W33A was found.}
    
\item {A point-like object offset by $\sim$3\arcsec to the SE of W33A was identified in the NIFS FOV. This object (Source\,B) has a relatively blue spectrum, which is consistent with the bright $J$-Band source identified in the $JHK$ maps. Source B is also associated with a $H$-band extended emission oriented to the NW direction that overlaps the emission from W33A, contributing to the observed $K$-band continuum extended emission observed in the NIFS FOV.}
    
\item {A third point-like source (Source C), offset by 6\arcsec to the S direction of W33A, was identified in the large-scale near-infrared maps. Source C is likely the driving source of a set of \hh knots oriented in the E-W direction. The identification of sources B and C revealed that W33A is located in a region with multiple (proto)stellar objects.}
\end{enumerate}

We reconstructed the data cube using only the first five Principal Components which represents 99.996\% of the variance and from this data cube we derive the following conclusions:

\begin{enumerate}
  \setcounter{enumi}{5}
\item First, we derived the physical parameters of the \hh jet and found that it is associated with an excitation temperature of (2.2\,$\pm$\,0.1)\,$\cdot$\,10$^3$\,K, and a mean \hh column density of $\log$(\ntot/cm$^{-2}$)\,=\,18.7$\pm$0.2. The mean $K$-band extinction is about 4.1$\pm$0.2\,mag (corresponding to an $A_{\rm V}$ of $\sim$61\,mag), and the mass of the \hh jet is about $\log$(M$_{H2}$/\msun)\,=\,$-$4.3$\pm$0.3.

\item We reanalysed the spectro-astrometry of the \brg emission, confirming that the ionised gas is tracing the cavity of the jets at sub-pixel scales. The positional offset between the blue- and red-shifted components is $\sim$0\farcs55, about three times smaller than previously reported ($\sim$1\farcs5), corresponding to a linear size of $\sim$1.3\,AU at 2.4\,kpc.

\item We measured the kinematics of the CO absorption lines to infer the dynamical mass of the W33A protostar. The analysis of the position-velocity diagram of the CO absorption features indicates a truncated rotating structure, with Keplerian-like rotation in the inner 0\farcs2 radii. We used a Keplerian model to evaluate the mass of the ``protostar + disc'' system as 10.0$^{+4.1}_{-2.2}$\,\msun assuming a distance of 2.4\,kpc. The mass is about $\sim$33\% smaller than previously reported by \citet{Davies10}, {placing W33A closer to the lower mass-limit for a high-mass protostar.}

\item The above-mentioned results were obtained thanks to the application of the PCA Tomography towards a \textit{bona fide} high-mass protostar, showing the technique is encouraging to be applied to other high-mass young stellar object candidates.

\end{enumerate}

\section*{Acknowledgements}

FN thanks to Funda\c{c}\~{a}o de Amparo \`a Pesquisa do Estado de S\~{a}o Paulo - FAPESP for  support through proc. 2017/18191-8. AD acknowledges FAPESP for support through proc. 2019/02029-2.
The authors would like to thank the valuable comments and suggestions made by the anonymous referee that helped to improve the manuscript.

\section*{Data availability}

The data sets were derived from observations available in the public domain: \url{http://www1.cadc-ccda.hia-iha.nrc-cnrc.gc.ca/gsa/} (under the Gemini program GN-2008A-Q-44).

\bibliographystyle{mnras}


\appendix
\section{Aperture photometry of the UKIDSS JHK images}
\label{appendix_phot}

The $JHK$ magnitudes of the W33A, Source\,B and Source\,C are listed in Table\,\ref{tab_photometry}.
The full list containing the 287 sources is available online as supplementary material and in the CDS.

\setlength{\tabcolsep}{3.5pt}
\begin{table*}
\caption{$JHK$ Photometry of the sources (first 10 rows). The full table is available online as supplementary material and in the CDS.}
\label{tab_photometry_all}
\centering
\begin{tabular}{lrr|c|rrrrrr|rrrrrr}
\hline
\hline
ID  & RA        & Dec       & UKIDSS  & J     & $\sigma_J$ & H     & $\sigma_H$   & K     & $\sigma_K$   & $J-H$  & $\sigma_{J-H}$   & $J-K$   & $\sigma_{J-K}$   & $H-K$   & $\sigma_{H-K}$   \\
  & (J2000)        & (J2000)       & Designation  & (mag)     & (mag)  & (mag)     & (mag)  & (mag)     & (mag)  & (mag)     & (mag)  & (mag)     & (mag)  & (mag)     & (mag)   \\
\hline
1$^a$   & 273.66458 & -17.86669 & J181439.51-175200.9 & 20.77$^\ast$ & --  & 15.05$^\ast$ & -- & 9.51  & 0.19 & 5.72 & 0.64  & 11.26 & 0.66  & 5.54  & 0.21  \\
2$^b$   & 273.66486 & -17.86736 & J181439.58-175202.6 & 15.45$^+$ & 0.08  & 13.76 & 0.09 & 10.30 & 0.20 & 1.70 & 0.13  & 5.16  & 0.22  & 3.46  & 0.22  \\
3$^c$   & 273.66495 & -17.86827 & --            & 20.77$^\ast$ & --  & 16.31 & 0.11 & 12.69 & 0.23 & 4.46 & 0.64  & 8.08  & 0.68  & 3.62  & 0.26  \\
4   & 273.66562 & -17.86503 & J181439.75-175154.1 & 11.37 & 0.07  & 10.99 & 0.08 & 10.29 & 0.20 & 0.38 & 0.11  & 1.08  & 0.21  & 0.70  & 0.21  \\
5   & 273.66211 & -17.85791 & J181438.90-175128.3 & 13.24 & 0.08  & 12.63 & 0.09 & 12.35 & 0.21 & 0.61 & 0.12  & 0.89  & 0.22  & 0.28  & 0.23  \\
6   & 273.67337 & -17.87352 & J181441.61-175224.6 & 13.35 & 0.08  & 13.07 & 0.09 & 12.93 & 0.21 & 0.28 & 0.12  & 0.42  & 0.23  & 0.14  & 0.23  \\
7   & 273.67474 & -17.86697 & J181441.94-175201.1 & 13.57 & 0.08  & 13.18 & 0.09 & 12.99 & 0.21 & 0.39 & 0.12  & 0.59  & 0.23  & 0.20  & 0.23  \\
8   & 273.66714 & -17.88072 & J181440.11-175250.5 & 14.11 & 0.08  & 13.14 & 0.09 & 12.63 & 0.21 & 0.96 & 0.12  & 1.48  & 0.22  & 0.51  & 0.23  \\
9   & 273.66238 & -17.87348 & J181438.97-175224.5 & 14.10 & 0.08  & 13.30 & 0.09 & 12.88 & 0.21 & 0.81 & 0.12  & 1.22  & 0.23  & 0.42  & 0.23  \\
10  & 273.66678 & -17.86504 & J181440.03-175154.1 & 14.29 & 0.08  & 13.40 & 0.09 & 12.95 & 0.21 & 0.89 & 0.12  & 1.35  & 0.23  & 0.46  & 0.23  \\
\hline
\end{tabular} \\
\noindent {\textbf{Notes:} the columns are as follows:
(1) ID of the source ($^a$: W33A; $^b$: Source\,B; $^c$: Source\,C);
(2) Right ascension (J2000, in degrees);
(3) Declination (J2000, in degrees);
(4) Name of the source or its designation in the UKIDSS catalogue;
(5) $J$-band magnitude;
(6) $J$-band magnitude error;
(7) $H$-band magnitude;
(8) $H$-band magnitude error;
(9) $K$-band magnitude;
(10) $K$-band magnitude error;
(11) $J-H$ colour index (in mag);
(12) $J-H$ colour index error (in mag);
(13) $J-K$ colour index (in mag);
(14) $J-K$ colour index error (in mag);
(15) $H-K$ colour index (in mag);
(16) $H-K$ colour index error (in mag).
An $\ast$ symbol indicates upper limit due to the non-detection of a point-like source at the given position in the corresponding filter.
$+$ indicates the $J$-band flux of a foreground source lying very close to the targeted object.}
\end{table*}
\setlength{\tabcolsep}{6pt}

\bsp	
\label{lastpage}
\end{document}